# Data-driven local operator finding for reduced-order modelling of plasma systems: II. Application to parametric dynamics

F. Faraji*[1], M. Reza*, A. Knoll*, J. N. Kutz**

* Plasma Propulsion Laboratory, Department of Aeronautics, Imperial College London, London, United Kingdom

** Department of Applied Mathematics and Electrical and Computer Engineering, University of Washington, Seattle, United States

**Abstract**: The dynamics of many real-world systems is dependent on certain parameters. These are either intrinsic to the system, inherently evolving over time, or that are external and enable controlling the system's behavior. A fully generalizable model shall be able to reliably represent these parametric behaviors. Plasma technologies serve as an example of parametric dynamical systems. For instance, in a Hall thruster – an industrially important plasma propulsion technology for spacecrafts – the dominant plasma phenomena governing the device's global dynamics change as the "self-sustained electric field" parameter varies over one characteristic cycle of the system. In addition, changing the control parameters, such as the intensity of externally applied magnetic field or the applied discharge voltage can majorly alter the dynamics of the thruster. In this Part II, we demonstrate that our novel data-driven local operator finding algorithm, Phi Method, which was discussed in Part I, can effectively learn the parametric dynamics so as to faithfully predict the systems' behavior over unseen parameter spaces. We describe two adaptations of Phi Method toward parametric dynamics discovery, namely, the "parametric Phi Method" and the "ensemble Phi Method". Two demonstration cases are adopted: one, the 2D problem of a fluid flow past a cylinder, and the other, the 1D problem of Hall thruster plasma discharge. For the first test case, the "parametric Phi Method" is assessed comparatively against the parametric implementation of OPT-DMD. The predictive performance of the parametric Phi Method notably surpassed that of the "parametric OPT-DMD" in the fluid test case. Across both test cases, the parametric and ensemble Phi Method were rather equivalently well able to recover the governing parametric PDEs and provide accurate predictions over the test parameters. Analysis of the ensemble ROMs underlined that Phi Method learns the coefficients of the dominant terms in the dynamics with very high confidence.

**Section 1: Introduction**

The high-fidelity numerical study and prediction of complex systems, in particular plasmas, have continued to persist among the grand challenges of the modern science and engineering. Plasmas, marked by their multifaceted and multiscale dynamics and inherent nonlinearities, require advanced modeling techniques for faithful representation of their behavior and achievement of predictive capabilities. However, conventional modeling approaches often face great computational challenges when employed for the analysis of high-dimensional systems like plasmas [1]-[7]. The advancements in data-driven (DD) reduced-order modeling techniques in recent years offer a promising pathway to tackle the above complexities by leveraging the wealth of information contained within the ever-increasing numerical and experimental datasets.

The data-driven and machine-learning (ML) methods enable the extraction of meaningful patterns and reduced representations of the involved physics from the vast and high dimensional datasets. By capturing the essential dynamics and dominant structures, data-driven reduced-order models (ROMs) can not only circumvent many of the computational complexities associated with conventional methods but can also enhance the interpretability of the underlying physics. Additionally, reliable ROMs can contribute to efficient planning and execution of the experiments by identifying the optimal experimental designs [8]. The DD ROMs can also facilitate the optimization of plasma technologies, streamline their development cycles, and underpin real-time analysis and control of plasma technologies (through the overarching, transformative concept of digital twins [9][10]).

Like most dynamical systems, plasmas can be described by a set of partial differential equations (PDEs). The plasma PDEs capture the evolution of key physical quantities, such as the densities, velocities, and temperatures of the plasma species (electrons and ions) as well as possibly the electromagnetic fields. A key challenge in accurately describing the plasma dynamics through PDEs arises from the inherent variability in plasma behavior under different conditions and operating regimes, which serve as additional parameters on top of the system's state variables. Parametric PDEs address this by introducing parameters to represent diverse conditions. These parameters can represent physical or geometric properties, the environmental factors, and/or the external control inputs. Particularly in plasma devices, the important parameters can be the operating conditions such as the applied

---

voltage, and the magnitude and topology of the externally applied magnetic field, or the intrinsic state variables like the average operating density and temperature, the ions' type (hence, their mass, ionization energy, and collision frequencies), and the material properties of the surfaces of the device that are in contact with the plasma.

Despite the near real-time capabilities of most DD ROMs to generate forecasts, their often-limited adaptability to parametric variations presents a significant challenge to achieving broader generalizability for the models. As a result, parameter embedding emerges as a crucial consideration in data-driven ROMs. Otherwise, when faced with variation in parameters, the model needs to be rebuilt through training on the data related to the new parameter, which serves as a major drawback for any practical reduced-order models intended for real-world applications.

In this article, we present two distinct data-driven approaches for parametric reduced-order modeling. The first and the primary approach is the parametric Phi Method, which belongs to PDE-discovery techniques. These techniques work by applying ML/DD methods, for instance neural networks [11][12] or sparse regression [13][14], to identify the governing PDEs directly from data. In addition to parametric Phi Method, we also explore a second DD approach, i.e., the parametric DMD, which falls into the POD-based reduced-order modeling category [15] and involves the global modes identification [16].

In the PDE-discovery techniques, the parameters are directly incorporated into the model. This explicit treatment of parameters allows PDE-based models to naturally adapt to changes in parameters. It also provides a more direct and generalizable representation of how the system may respond to different parameter values. On the contrary, whereas POD-based models have proven successful in capturing dominant modes in the data and reducing the data dimensionality [17]-[21], they may not be the optimal choice when the focus would be on explicitly modeling parametric dependencies. Indeed, in extending the DMD algorithm to accommodate parametric variations, a common approach entails employing interpolation techniques between the modes associated with the neighboring parameter values [22] (details in Section 2.1). Nonetheless, the interpolation between the modes assumes a smooth transition in the parameter space among the different parameter values. Accordingly, in the cases where a system's behavior undergoes significant changes across the parameter space, this approach will become impractical because of the substantial training dataset that will be needed to adequately sample the parameter space.

In Part I of the present two-part article [23], we focused on introducing the Phi Method algorithm. We verified the resulting ROMs from Phi Method for systems that featured fixed set of parameters, comparing the models' performance against the performance of the ROMs from the Optimized Dynamic Mode Decomposition (OPT-DMD) method [24]. Here, our objectives are: first, to assess the capability of Phi Method in capturing the parametric dependencies in systems' dynamics and, second, to evaluate the generalizability of the discovered Phi Method models across a wide range of system parameters, comparing again, where relevant, the Phi-Method-derived ROMs against those from OPT-DMD.

Accordingly, we begin by presenting the parametric extensions to the formulations underlying Phi Method and OPT-DMD, obtaining algorithms that we refer to as "parametric Phi Method" and "parametric OPT-DMD", respectively. Next, for the purpose of the demonstrations in this part II, we use two test cases: (1) a 2D flow around a cylinder at various Reynolds number conditions, (2) a 1D azimuthal Hall-thruster-representative E × B plasma discharge, where the plasma is immersed in a perpendicular configuration of externally applied electric and magnetic fields, the intensities of which span over a range of values.

### Section 2: Description of the methods

In a parametric system, the dynamics is not only dependent on the state variables of the system but is also influenced by some additional parameters. The inclusion of these additional parameters is necessary to obtain a comprehensive and adaptable representation of complex systems that could be applicable to a broad range of scenarios. In this section, we present parametric extensions to the two data-driven algorithms that we, respectively, reviewed and introduced in Part I paper [23], i.e., the DMD algorithm, and the Phi Method algorithm. Following the same order of methods' description as that pursued in Part I [23], we first present the "parametric DMD (OPT-DMD)" followed by the "parametric Phi Method".

The parametric implementations of DMD and Phi Method enable the development of ROMs for dynamical systems that exhibit parametric dependencies. Such dynamical systems are described by the following equation

$$\frac{d\boldsymbol{x}(t)}{dt} = \mathcal{F}(\boldsymbol{x}(t); \boldsymbol{\mu}). \tag{Eq. 1}$$

In Eq. 1, $x$ is the state vector and $\mu$ is the "parameter vector" containing the individual parameters ($\mu_j$ with $j = 1, 2, \ldots n_p$) that the system's dynamics is dependent on, hence,

$$\mu = \left[\mu_1, \mu_2, \ldots, \mu_{n_p}\right]^T. \tag{Eq. 2}$$

## 2.1. Parametric Dynamic Mode Decomposition (DMD)

The parametric DMD is an extension to the DMD algorithm presented in Part I [23] of this article. The method is intended for reduced-order modelling of dynamical systems with one or more parametric dependencies. Multiple variants of parametric DMD are proposed in the literature [22][25], which rely on interpolation to achieve parameter realization. The proposed algorithms differ in the way that they interpolate between the parameters. In the following, we describe the specific parametric DMD algorithm that we have used in this study.

We start by gathering the data snapshots, $X^{\mu_s}$, that correspond to different sets of values of the parameter vector ($\mu$) for the parametric dynamical system described by Eq. 1. These different sets of values are denoted as $\mu_s \in M$ ($s = 1, 2, \ldots, S$), where

$$M = \{\mu_1, \mu_2, \ldots, \mu_S\}. \tag{Eq. 3}$$

The arrangement of the spatiotemporal data in matrix $X^{\mu_s}$ is identical to that in the basic DMD method [23]. Hence, $X^{\mu_s}$ comprises the time series of column-wise rearranged snapshots $X^{\mu_s} \in R^{n \times m-1}$, where $n$ represents the dimension of each data snapshot, and $m$ specifies the total number of snapshots.

We apply the OPT-DMD algorithm [24] to the individual dataset $X^{\mu_s}$ to determine the eigenvalues ($\Lambda^{\mu_s}$) and the eigenfunctions ($\Psi^{\mu_s}$) of the DMD's operator $A^{\mu_s}$ [23] for each parameter realization such that

$$X'^{\mu_s} \approx A^{\mu_s} X^{\mu_s} \tag{Eq. 4}$$

where, $X'^{\mu_s}$ represent the data matrix $X^{\mu_s}$ shifted by one time step forward.

The amplitudes of the modes corresponding to each parameter value set ($B^{\mu_s}$) are also directly obtained from the OPT-DMD algorithm. The matrices $\mathbf{\Lambda}$, $\mathbf{\Psi}$, and $\mathbf{B}$ contain the eigenvalues, the spatial modes, and the modes' amplitude, respectively, corresponding to each parameter realization

$$\mathbf{\Lambda} = \begin{bmatrix} \Lambda^{\mu_1} \\ \Lambda^{\mu_2} \\ \vdots \\ \Lambda^{\mu_S} \end{bmatrix}, \qquad \mathbf{\Psi} = \begin{bmatrix} \Psi^{\mu_1} \\ \Psi^{\mu_2} \\ \vdots \\ \Psi^{\mu_S} \end{bmatrix}, \qquad \mathbf{B} = \begin{bmatrix} B^{\mu_1} \\ B^{\mu_2} \\ \vdots \\ B^{\mu_S} \end{bmatrix}. \tag{Eq. 5}$$

For a new parameter vector $\mu_q \notin M$, the eigenvalues ($\Lambda^{\mu_q}$), the spatial modes ($\Psi^{\mu_q}$), and the corresponding amplitudes ($B^{\mu_q}$) are obtained by evaluating the Lagrangian interpolation between the respective quantities of the neighboring parameter vectors to $\mu_q$. Interpolation can be performed using linear, cubic, or any higher order Lagrange interpolant polynomials. Following the interpolation, the time series data for the "query" parameter vector $\mu_q$ can be reconstructed using the relation in Eq. 6

$$x^{\mu_q}|_{k+1} \approx \Psi^{\mu_q} B^{\mu_q} \Lambda^{\mu_q}|_k, \qquad k = 1, 2, \ldots. \tag{Eq. 6}$$

It is emphasized that the interpolation would result in a reasonable estimation of the dynamics only if the variation of the spatiotemporal modes is smooth across the parameter space. Strong nonlinear dependencies of the DMD modes to the systems' parameters can cause significant discrepancy between the interpolated and the "true" spatial and temporal characteristics of the modes for the query parameter-vector $\mu_q$.

Apart from using OPT-DMD in this work as the core DMD algorithm applied to the datasets of the individual parameters, our adopted scheme has minor differences with respect to the "reduced eigen-pair interpolation" (rEPI) approach that was followed by Huhn et al. in Ref. [25] for the interpolation among the spatial modes. In this regard, another noteworthy difference here is that, whereas we directly interpolate the DMD modes ($\Psi^\mu$) for a new parameter realization, in Ref. [25], the eigenvectors ($V^\mu$) of the DMD's operator and the spatial POD modes ($\widetilde{U}^\mu$) were interpolated separately. Subsequently, the DMD modes were computed as $\Psi^\mu = \widetilde{U}^\mu V^\mu$ [25].

In Ref. [25], another parametric DMD algorithm has also been presented, which is called the "reduced Koopman operator interpolation" (rKOI). This algorithm involves interpolating directly the reduced DMD operator $\tilde{A}^\mu$ and then calculating the $\Lambda^\mu$ and $\Psi^\mu$ directly from eigendecomposition of $\tilde{A}^\mu$.

Finally, a rather different parametrization technique is suggested in Ref. [27]. In this approach, all data matrices for the different parameter realizations ($X^{\mu_s}$) are stacked column-wise in an augmented data matrix $X \in R^{(n_p \times n) \times m-1}$. The DMD is then applied on the augmented data matrix $X$ [27]. A limitation of this approach is that, while the spatial modes and their amplitudes are specific to each parameter realization, the eigenvalues (time dynamics) are shared and assumed constant across the entire parameter space.

## 2.2. Parametric Phi Method

Recalling from Part I [23] of the present article, Phi Method discovers discretized partial differential equations describing a system's evolution ($\mathcal{F}$ in Eq. 1). This discretized dynamics is encapsulated within a linear operator $\Phi$ such that

$$x_i^{k+1} = F(y_i^k)\Phi. \tag{Eq. 7}$$

In Eq. 7, $x_i^{k+1}$ represent the state vector on node $i$ at time step $k+1$, $y_i^k$ encompasses the state vectors on node $i$ and on its neighboring nodes at time step $k$, and $F$ is the library of candidate terms that can include linear and nonlinear observables ($f_l(y_i)$, $l = 1, 2, \ldots L$) of the state vector variables. Eq. 7 is assumed to be valid for every node within the computational domain ($i = 2, 3, \ldots, N_g - 1$, for a 1D domain) and for all the times ($k = 1, 2, \ldots$).

To train a Phi Method model, we evaluate the $x_i^{k+1}$ and the $F(y_i^k)$ for all the nodal points $i$ and at all the time steps $k$ within the training interval and store the values in two matrices $X$ and $\Theta$, respectively [23]. In case the number of training data points is very large, we can consider a randomly sampled subset of the points in space and time for the model training. Matrix $\Phi$ will be then determined by performing a least-squares regression as per Eq. 8

$$\Phi = \arg\min_{\Phi} \|X - \Theta\Phi\|_2 \approx \Theta^\dagger X \tag{Eq. 8}$$

The parametric extension to Phi Method ("*parametric Phi Method*" implementation), which enables the discovery of the dynamical systems given by Eq. 1, is straightforward. The only modification to what was described above involves incorporating additional observables into the library of candidate terms ($F$) such that

$$x_i^{k+1} = F(y_i^k, \mu)\Phi_{\text{Param}}. \tag{Eq. 9}$$

The additional observables can be expressed as various functions $h_j(y_i, \mu)$, with $j = 1, 2, \ldots, J$, of both the parameters and the state variables of the system. Therefore,

$$F(y_i, \mu) = [f_1(y_i), f_2(y_i), \ldots, f_L(y_i), h_1(y_i, \mu), h_2(y_i, \mu), \ldots, h_J(y_i, \mu)]. \tag{Eq. 10}$$

Another approach to use Phi Method for reduced-order modelling of dynamical systems is the so-called "*ensemble Phi Method*" approach. Ensembling techniques are commonly used for the data-driven and the machine-learning-based ROM developments [28]-[30]. Ensembling is the practice of aggregating multiple individual models to achieve an improved performance. By leveraging the diversity among the individual learned models, the ensembling approach aims to mitigate overfitting, to enhance generalization, and to increase the models' robustness to varying data patterns.

In the ensemble Phi Method, instead of training a single model on the complete dataset that covers the whole parameter space, we train individual models on the data that correspond to each set of values of the parameter vector within our training set ($\mu_s$, Eq. 3). This leads to $S$ independent models, with $S$ being the total number of realizations (sets of values) of the parameter vector $\mu$ in the training dataset. To aggregate the resulting models, we compute the average over $S$ number of coefficient matrices, $\Phi_s$, $s = 1, 2, \ldots S$, associated with each model, i.e., $\Phi_{\text{Ens}} = \frac{1}{S}\sum_{s=1}^{S} \Phi_s$. The average coefficient matrix (operator) $\Phi_{\text{Ens}}$ represents the ensemble Phi Method model, which is then used for the predictions as per Eq. 9, with $\Phi_{\text{Param}}$ replaced with $\Phi_{\text{Ens}}$.

For the fluid flow test case, the ensemble Phi Method model is discussed in the Appendix section. For the plasma discharge test case, the parametric and the ensemble Phi Method models are compared in terms of their performance and characteristics within Section 3.2.2.

## Section 3: Results

The predictive performance of parametric ROMs from Phi Method and OPT-DMD are assessed in this section. The predictions of the Phi-Method-derived ROMs are compared against those of the OPT-DMD-derived ROMs in the 2D fluid flow test case only (subsection 3.1). This is because, for this test case, the variability of the problem's dynamics across the parameter space as represented by the Reynolds number was found to be sufficiently smooth so that the interpolation associated with the parametric OPT-DMD would produce meaningful results. The moderate parametric dependency of the dynamics observed for the fluid flow test case was, however, found to not be the case for the plasma discharge test case, where the dynamics of the system exhibited significant variations in the parameter space. Therefore, for the plasma test case (subsection 3.2), only the results from parametric Phi Method are presented and discussed.

### 3.1. Test case 1: 2D flow past a cylinder

*3.1.1. Description of the problem setup*

This first test case involves the evolution of a 2D fluid flow around a stationary cylinder. The flow evolution is characterized by the recurring shedding of vortices in the downstream wake of the cylinder. The vortex shedding phenomenon is strongly influenced by the Reynolds number of the flow [31] – beyond a certain Reynolds number, the flow becomes unstable as it passes over the cylinder and small perturbation of the flow leads to the wake instability and the periodic formation of the vortices. In the intermediate range of Reynolds numbers, the flow is laminar, whereas it transitions to a turbulent state once the Reynolds number exceeds a specific threshold [31]. Assuming incompressible and barotropic flow, the spatiotemporal variation of the vorticity field $\Omega(x, y, t)$ can be described by

$$\frac{\partial \Omega}{\partial t} + (\boldsymbol{V}.\boldsymbol{\nabla})\Omega = \frac{1}{Re}\nabla^2\Omega. \tag{Eq. 11}$$

Eq. 11 represents a parametric dynamical system with the parameter being the freestream's Reynolds number ($Re$).

The data for this parametric system were generated from several fluid simulations that we performed using the "ViscousFlow.jl" package [32] across an intermediate range of Reynolds numbers, where the flow exhibits the oscillatory vortex-shedding pattern while remaining laminar.

To briefly describe the simulations' setup, the computational domain is $6 \times 4$ cm 2D box discretized using $300 \times 200$ nodes. A cylinder of 1-cm diameter is included in the domain, with its center positioned 1 cm away from the left boundary. The free stream of the flow enters the domain at a specific pre-determined Reynolds number. Each simulation was run for a total duration of 100 s, with the flow data collected at 0.1 s intervals. Further details on the simulations' setup and conditions are provided in Part I of the paper [23].

The simulations were performed over a broad set of Reynolds numbers, namely, $Re \in \{100, 150, 175, 200, 250, 275, 300, 350, 400, 425, 450, 500\}$. All of the data corresponding to this broad range of Reynolds numbers were divided into the training and the testing sets in two fashions: one, $Re_{train} \in \{100, 200, 300, 400, 500\}$ and $Re_{test} \in \{150, 250, 350, 450\}$, and two, $Re_{train} \in \{100, 150, 200, 250, 300, 350, 400, 450, 500\}$ and $Re_{test} \in \{175, 275, 425\}$. The first division of the data is referred to as "**dataset 1**" or **DS1**. The second division, which represents finer increments in the parameter space for the training, is called "**dataset 2**" or **DS2**. In general, we have used DS1 for the training and testing of the ROMs in the fluid flow test case unless it is explicitly stated otherwise.

For either dataset, the initial 45 seconds of the simulations, representing the transient state of the fluid system [23], were excluded from either the training or the test data. This was to enable deriving reliable ROMs from the OPT-DMD application to the data. It is underlined that the POD-based methods, including OPT-DMD, work reliably only when applied to the (quasi) steady-state data from a system [20][21][23].

The parametric OPT-DMD was trained on the time-series snapshots of the vorticity field belonging to the training parameter set. We would highlight that, in the parametric DMD, determining the optimal number of ranks in the underlying Singular Value Decomposition (SVD) (or, equivalently, the number of DMD modes to retain) involves a compromise between two aspects: (i) the percentage of information captured in the resulting low-rank representation on the training data points, and (ii) the accuracy of the interpolation in the parameter space on the test data points. In this regard, retaining a higher number of DMD modes improved the closeness of the predictions of an individual OPT-DMD ROM to the original training data corresponding to each specific parameter. However, we observed that the quality of the interpolation between the DMD bases degraded as the number of DMD modes increased. Having identified the optimal trade-off between the above two effects, we retained 10 DMD modes for this test case example.

In the case of Phi Method, the library of terms (Θ) can include both linear and nonlinear observables of the system's state vector as well as the involved parameter(s), which is the $Re$ for the present test case. The library terms considered for the fluid flow test case are shown in Eq. 12. The impact of the size of the library and the inclusion of extra terms was thoroughly investigated in Part I [23]. Hence, we have focused here only on the minimum set of dynamical terms that are essential to properly represent the dynamics of fluid flow system. Accordingly, Eq. 9 becomes

$$\Omega_{i,j}|_{k+1} = \left[ \begin{bmatrix} \Omega_{i,j-1} \\ \Omega_{i-1,j} \\ \Omega_{i,j} \\ \Omega_{i+1,j} \\ \Omega_{i,j+1} \end{bmatrix}^T \begin{bmatrix} V_{x_{i,j-1}} \\ V_{x_{i-1,j}} \\ V_{x_{i,j}} \\ V_{x_{i+1,j}} \\ V_{x_{i,j+1}} \end{bmatrix}^T \begin{bmatrix} V_{y_{i,j-1}} \\ V_{y_{i-1,j}} \\ V_{y_{i,j}} \\ V_{y_{i+1,j}} \\ V_{y_{i,j+1}} \end{bmatrix}^T \begin{bmatrix} (\Omega V_x)_{i,j-1} \\ (\Omega V_x)_{i-1,j} \\ (\Omega V_x)_{i,j} \\ (\Omega V_x)_{i+1,j} \\ (\Omega V_x)_{i,j+1} \end{bmatrix}^T \begin{bmatrix} (\Omega V_y)_{i,j-1} \\ (\Omega V_y)_{i-1,j} \\ (\Omega V_y)_{i,j} \\ (\Omega V_y)_{i+1,j} \\ (\Omega V_y)_{i,j+1} \end{bmatrix}^T \begin{bmatrix} \frac{1}{Re}\Omega_{i,j-1} \\ \frac{1}{Re}\Omega_{i-1,j} \\ \frac{1}{Re}\Omega_{i,j} \\ \frac{1}{Re}\Omega_{i+1,j} \\ \frac{1}{Re}\Omega_{i,j+1} \end{bmatrix}^T \right]_k \Phi_{\text{Param}}. \quad \text{(Eq. 12)}$$

Before training the Phi Method models, each flow property was normalized with respect to its spatiotemporally maximum value over the *entire* dataset that corresponds to all the simulated parameters. The left-hand side of Eq. 12 as well as the terms in the library were then evaluated on various nodes ($i = 2, 3, \ldots, N_i - 1, j = 2, 3, \ldots, N_j - 1$) and time steps ($k = 1, 2, \ldots m - 1$) using the normalized data across the training parameter space. $\frac{1}{Re}$, included in the last term within the library, captures the parametric dependency in the resulting PDE that describes the evolution of the vorticity. The optimal matrix of coefficients $\Phi_{\text{Param}}$ associated with the parametric Phi Method ROM was obtained by performing a regression on both sides of Eq. 12.

We also derived an ensemble Phi Method ROM for the fluid flow test case, which is mainly discussed in the Appendix section. For the ensemble Phi Method ROM, the individual models were trained separately on the simulation data corresponding to each Reynolds number of dataset 2 (DS2), i.e., $Re_{train} \in \{100, 150, 200, 250, 300, 350, 400, 450, 500\}$. This resulted in $S = 9$ independent models. Taking an average over the coefficient matrices of the individual models, the coefficient matrix (operator), $\Phi_{\text{Ens}}$, of the ensemble ROM was obtained. The library terms used to drive the ensemble ROM were the same as those shown in Eq. 12.

*3.1.2. Results*

The predictions of the trained parametric ROMs on the test sets of the $Re$ parameter are presented and discussed in this subsection.

Figure 1 shows the forecasts of the parametric Phi Method and the parametric OPT-DMD ROMs over the test $Re$ parameters of DS1. The results are presented in terms of the temporal evolution of the spatially averaged vorticity ($\Omega_{mean}$) and the local value of the vorticity field ($\Omega_{mid}$) in comparison to the ground-truth data from the simulations. The local vorticity value corresponds to the value at the mid-location of the domain ($x = 3$ cm and $y = 2$ cm). Figure 1(right column) additionally presents the time variations of the ROMs' loss factor. The loss factor quantifies the extent of error (or deviation) in the models' predictions relative to the true data. The definition of the loss factor and how to compute it were presented in Part I [23].

The plots of $\Omega_{mid}$ (Figure 1(middle column)) evidently show that the predictions of the parametric Phi Method ROM closely align with the true data. In contrast, the predictions of the parametric OPT-DMD ROM exhibit a relatively high level of discrepancy. According to Figure 1(right column), the loss associated with the forecasts of the parametric Phi Method ROM is, on average, about four times smaller than that of the parametric DMD ROM across the test parameters.

The fully reconstructed 2D snapshots of the normalized vorticity field from the parametric Phi Method and the parametric OPT-DMD ROMs at three different time instants are illustrated in Figure 2 for the test Reynolds numbers of DS1. Furthermore, in Figure 3, the absolute-difference maps between the predicted snapshots and the respective "true" snapshots are plotted.

The close resemblance between the Phi Method ROM's snapshots and the ground-truth further demonstrates the ability of Phi Method to reliably capture the parametric dependency in the dynamics and, hence, to provide successful predictions of the desired quantity across the parameter space. Contrarily, the parametric OPT-DMD does a rather unsatisfactory job in embedding the parametrization underlying the dynamics. This is evident from the relatively significant discrepancies observed between its predicted snapshots and the ground-truth. In this respect, although we noticed in Part I [23] that the OPT-DMD exhibited a slightly better performance compared to Phi Method toward time forecasting of the fluid flow system at a fixed parameter value, the present results from Figure 2 and Figure 3 underline that the parametric Phi Method provides substantially more accurate predictions when the defining parameter of a system undergoes variations.

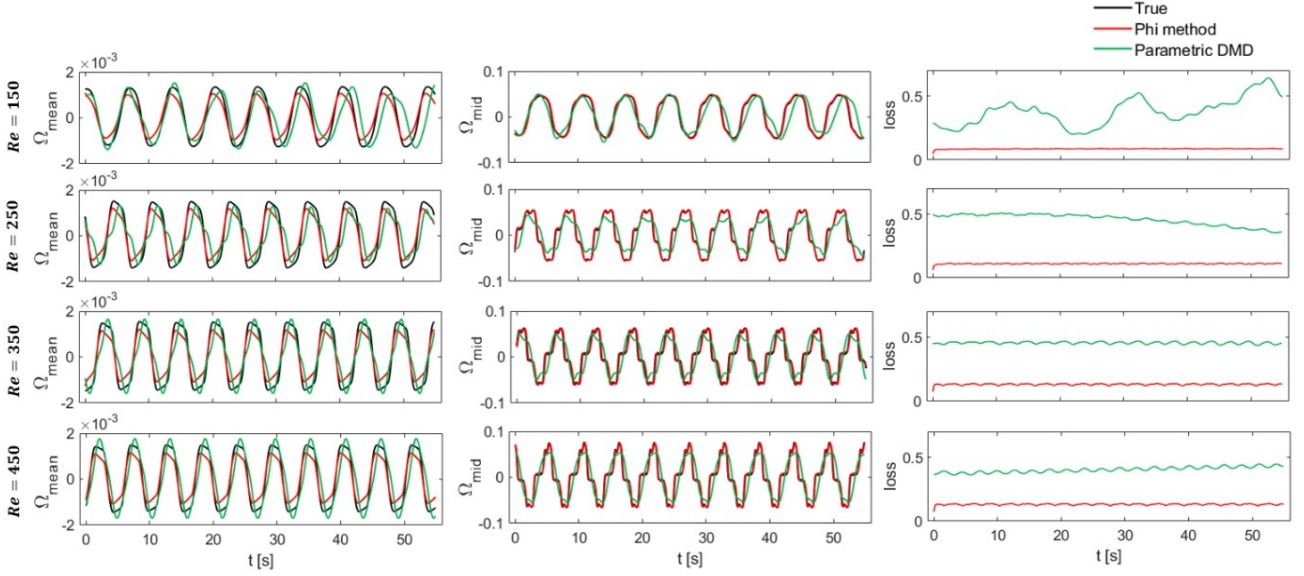

Figure 1: Comparison of the predictions from the parametric Phi Method and the parametric OPT-DMD ROMs against the ground-truth data for the test case 1 across the test Reynolds numbers belonging to DS1; time evolutions of the spatially averaged normalized vorticity (**left column**), and local values of the normalized vorticity at the mid-location within the simulation domain (**middle column**). (**right column**) Time evolutions of the loss factor calculated over the entire domain.

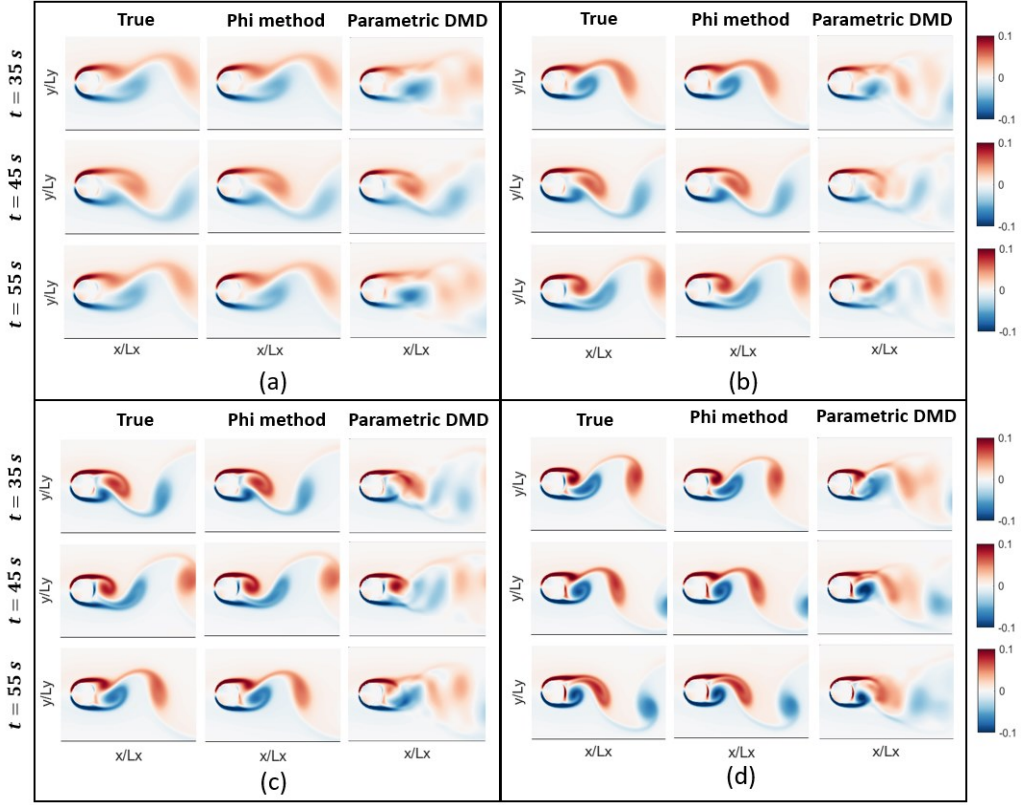

Figure 2: Comparison of the predictions from the parametric Phi Method and the parametric OPT-DMD ROMs against the ground-truth data for the test case 1 across the test Reynolds numbers belonging to DS1; 2D snapshots of the normalized vorticity field at three different sample time instants for the test parameters of (a) $Re = 150$, (b) $Re = 250$, (c) $Re = 350$, and (d) $Re = 450$.

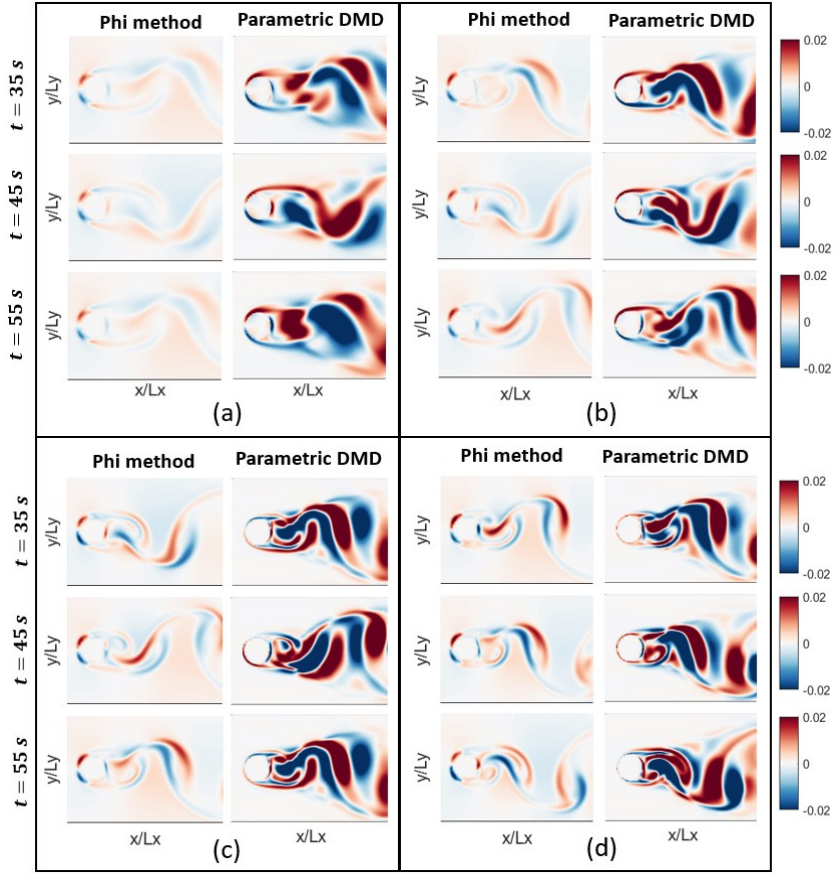

Figure 3: Comparison of the predictions from the parametric Phi Method and the parametric OPT-DMD ROMs for the test case 1; absolute-difference plots between the true and the predicted normalized vorticity snapshots at three time instants for the test parameters of DS1, namely, (a) *Re* = 150, (b) *Re* = 250, (c) *Re* = 350, and (d) *Re* = 450.

We would emphasize that the "poor" performance of the parametric OPT-DMD in the fluid flow test case is not a limitation of the OPT-DMD itself as a data-driven method, but rather it is related to the limited accuracy of the interpolation between the DMD bases. In this regard, it should be noted that the optimality of the DMD modes is not preserved through the interpolation. In other words, the resulting interpolated modes may no longer represent the best low-rank approximation of the system that satisfies the linearity of time dynamics.

When the variation of the DMD modes in the parameter space occurs on a non-smooth and strongly nonlinear manifold, or when the two parameters between which we are interpolating the DMD modes are too far apart, the interpolated modes can exhibit a larger extent of deviation from their optimality condition. This argument is further supported by the results provided in Appendix A, where we present the outcomes of the parametric OPT-DMD model when trained on the expanded dataset 2 (DS2). The results in Appendix A show a significant improvement in the accuracy of the parametric OPT-DMD ROM with respect to the results presented so far (from training on DS1).

Besides expanding the dataset, employing more advanced interpolation schemes may also improve the quality of the parametric OPT-DMD ROM by allowing for the interpolation to be reliably performed between more widely spaced parameters. These advanced interpolation schemes include techniques that utilize an intermediary linear subspace called "Grassman manifold" to perform the interpolation and then transfer back to the subspace of the POD modes [33].

In Figure 4, we have compared the true and the interpolated spatial modes corresponding to the first four leading DMD bases across the test *Re* parameters of DS1. The true bases (modes) are obtained by applying OPT-DMD directly to the data associated with each specific test Reynolds number. It is evident from the plots in Figure 4 that, in some cases, the interpolated modes are significantly different from their true counterparts. As we have shown in Appendix A, this discrepancy is reduced when the model is trained on DS2, for which the interpolation is carried out between DMD modes that correspond to the parameters in closer proximity within the parameter space.

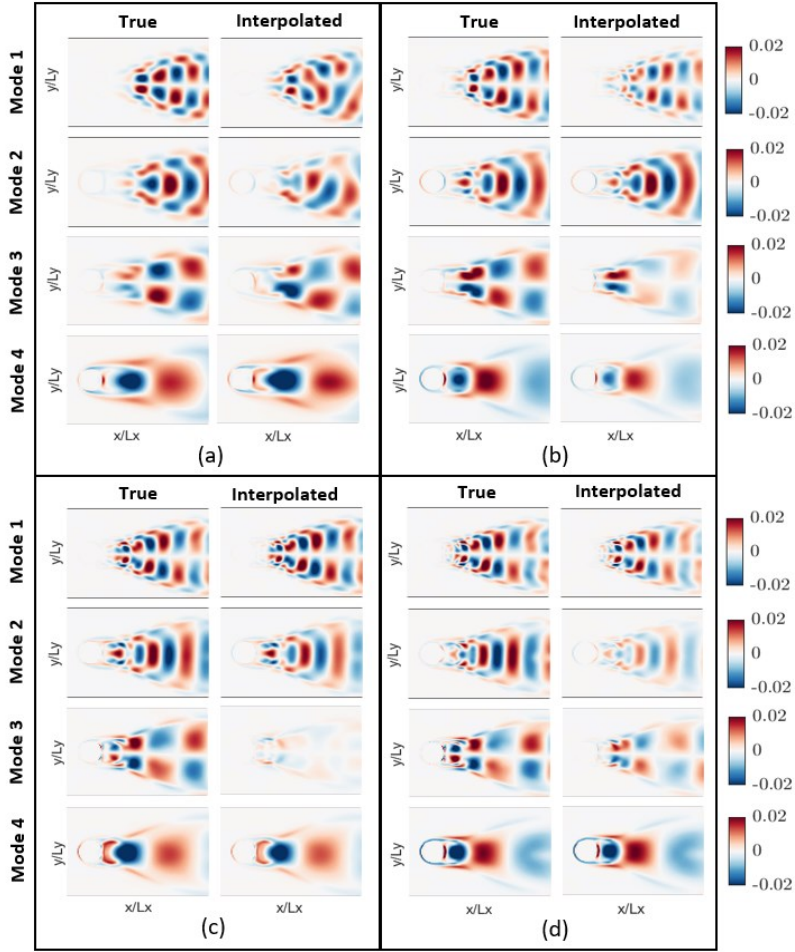

Figure 4: Visualization of the first four interpolated DMD modes associated with the parametric OPT-DMD for the test case 1 over the test Reynolds numbers of DS1. The interpolated modes are compared against the corresponding true DMD modes. (a) $Re = 150$, (b) $Re = 250$, (c) $Re = 350$, and (d) $Re = 450$.

In Figure 5, the spatiotemporally averaged losses of the predictions from the parametric Phi Method and the parametric OPT-DMD ROMs with respect to the ground-truth data of the fluid flow test case are plotted for various Reynolds numbers. The results are presented for the ROMs trained on DS1 and on DS2.

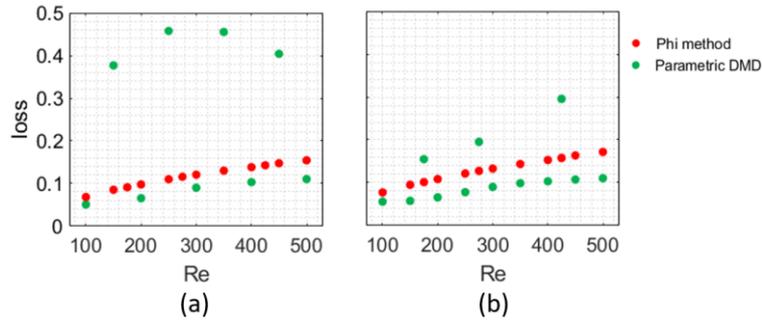

Figure 5: Comparison of the predictions from the parametric Phi Method and the parametric OPT-DMD ROMs for the test case 1; plots of the time-averaged (mean) values of the loss factor vs. the Reynolds number from the ROMs trained on (a) dataset 1 (DS1), and (b) dataset 2 (DS2). The loss factor is calculated over the entire domain.

Referring to Figure 5, the spatiotemporally averaged loss of the Phi Method ROM is seen to remain identical between the training on DS1 and DS2 across the range of $Re$ values. Additionally, for Phi Method, the ROM's loss level remains similar among the training and the test data. However, in the case of the parametric OPT-DMD, the loss in predicting the test data is considerably higher than the loss associated with the reconstruction of the training data. Moreover, it is again observed that the number of samples of the parameter space that are included in the training greatly influences the disparity level between the loss of the parametric OPT-DMD ROM on the training and on the test data.

The difference between the accuracy of the parametric Phi Method ROM and the parametric DMD ROM lies in a fundamental difference between these two techniques. In fact, by identifying the local relationships and correlations within the data, Phi Method approximates the underlying PDE describing the data. Consequently, this method, like any other data-driven PDE-discovery technique, enables the explicit incorporation of the parametric dependencies present in the system into the model. In contrast, POD-based approaches, such as DMD, do not explicitly account for the parametric dependencies in their formulation. They identify the modes based on the covariance of the dataset without differentiating between the different parameter values. Accordingly, for the parametric DMD, although a distinct set of modes is associated with each parameter, there is no explicitly quantified relationship between the specific parameter regimes and their corresponding modes. This "implicit" treatment in parametric DMD limits the model's ability to capture the intricate relationships between the parameters and the system's response.

As a final discussion, we have illustrated in Figure 6 the rectangular rearrangement of the matrices of coefficients ($\Phi_{Param} \in R^{30\times1}$, or $\Phi_{Param} \in R^{5\times6}$ after rearrangement) for the parametric Phi Method ROM when trained on DS1 and on DS2. It is noticed that the matrices (operators) of the two models are almost identical.

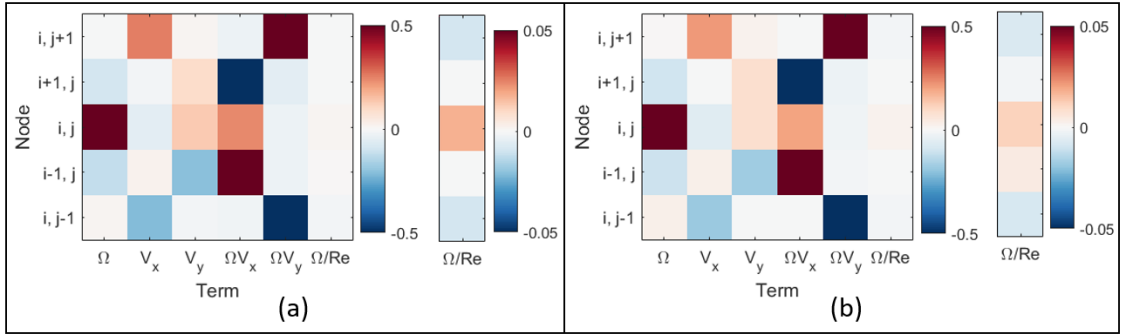

Figure 6: Rearranged normalized representations of the coefficients matrices ($\Phi_{Param}$s) of the parametric Phi Method ROM for the test case 1; (a) from training on DS1, and (b) from training on DS2.

The last column of the matrices in Figure 6 captures the parametric dependency of the dynamics to the Reynolds number, as was represented by the right-hand-side term in the theoretical relation describing the dynamics of the vorticity (Eq. 11). Discretizing this term in Eq. 11 using a finite difference scheme, we obtain

$$\frac{1}{Re}\nabla^2\Omega = \frac{1}{\Delta x^2}\left(\frac{\Omega_{i+1,j}^k}{Re} - \frac{2\Omega_{i,j}^k}{Re} + \frac{\Omega_{i-1,j}^k}{Re}\right) + \frac{1}{\Delta y^2}\left(\frac{\Omega_{i,j+1}^k}{Re} - \frac{2\Omega_{i,j}^k}{Re} + \frac{\Omega_{i,j-1}^k}{Re}\right) \quad \text{(Eq. 13)}$$

where, $\Delta x$ and $\Delta y$ denote the cell sizes of the grid discretizing the domain along the $x$ and $y$ directions, respectively.

Given that $\Delta x = \Delta y$ for the data used to train the ROMs, we could expect a nearly uniform distribution of coefficients for the $\frac{\Omega}{Re}$ term across the surrounding nodes of the node $(i,j)$. However, looking at Figure 6, it is interesting that this is not the case. In fact, the magnitude of the coefficients for $\frac{\Omega_{i,j-1}^k}{Re}$ and $\frac{\Omega_{i,j+1}^k}{Re}$ are larger than those corresponding to $\frac{\Omega_{i-1,j}^k}{Re}$ and $\frac{\Omega_{i+1,j}^k}{Re}$. This disagreement between This disagreement between the coefficients matrices of the two Phi Method ROMs visualized in Figure 6 and what is prescribed by the finite differencing can imply that the identified stencil by Phi Method is an alternative optimal stencil for the discretization of the governing PDE. This must, of course, be considered in conjunction with the coefficients of the other library terms and their associated discretization stencils.

We also point out that the rest of the columns of the visualized matrices in Figure 6 (those that do not reflect the parametric dependence on the $Re$) are closely aligned with the $\Phi$ matrices obtained from application of "basic" Phi Method to a single-parameter dataset as was discussed in Part I [23].

In Appendix B, we have presented the $\Phi_{Ens}$ matrix corresponding to the ensemble Phi Method ROM for the fluid flow test case. It is recalled from Section 3.1.1 that the $\Phi_{Ens}$ has been derived by averaging individual $\Phi$ matrices that were obtained by training Phi Method on single datasets associated with the individual Reynolds number values. In Appendix B, we have also shown the standard deviations of the coefficients of the library terms across the individual models. The mean $\Phi$ matrix in the ensemble model ($\Phi_{Ens}$) is particularly found to be very similar to the matrices shown in Figure 6 in terms of the coefficients learned for the library terms. This means that, for the fluid flow test case, the ensemble ROM is rather equivalent to the "standard" parametric ROM.

**3.2. Test case 2: 1D azimuthal plasma configuration**

*3.2.1. Description of the problem setup*

This test problem represents a 1D azimuthal geometry of a Hall thruster – a spacecraft plasma propulsion technology. The plasma in this test case evolves in the presence of externally applied, mutually perpendicular electric ($E$) and magnetic ($B$) fields. The E × B (cross-field) plasma configurations, similar to that of a Hall thruster, are of remarkable scientific interest across various plasma science domains. This is largely because of the rich underlying physics of the E × B plasmas that is characterized by a broad spectrum of instabilities and turbulence that interact with the plasma species. In particular, a long-standing universal question in the physics of cross-field plasmas regards the "anomalous" transport of the electrons across the magnetic field. Anomalous transport refers to the enhanced transport of the plasma particles compared to what is predicted by the classical theories. In recent years, numerous studies have pointed to the significant contributions of plasma instabilities to the anomalous transport of the electrons [34][35]. However, a universally applicable model to predict the instabilities and their induced transport is yet to be achieved [2][4][36].

The adopted 1D setup serves as a simplified but representative geometry that captures instabilities and fluctuations as well as the interactions of these with the plasma particles. The characteristics of the instabilities and their impacts on the plasma vary with the magnitudes of the applied electromagnetic field. As a result, the applied axial electric field ($E_x$) and the radial magnetic field ($B_y$) act as the parameters of the dynamical system that represents this plasma discharge configuration. Hence, this problem provides an ideal plasma test case to demonstrate that the Phi Method's ROM can capture these parametrizations and to also illustrate the generalizability of the derived relationship(s) from Phi Method to a broad range of parameters.

IPPL-1D particle-in-cell (PIC) code [37][38] is used to simulate the plasma in this adopted test case and so to generate the dataset for the various values of $E_x$ and $B_y$. The simulations' setup follows what was described in Part I [23] except for a few modifications. An overview of the setup here is provided below.

The simulation domain is a Cartesian section along the azimuthal ($z$) direction of a Hall-thruster-representative geometry. The $z$-direction is perpendicular to the directions of the applied electric field, that is along the axial coordinate ($x$), and the applied magnetic field, that is along the radial coordinate ($y$).

To ensure that the domain size remains larger than the characteristic length of the instabilities across the entire simulated range of $E_x$ and $B_y$ parameters, the azimuthal extent of the domain is increased to 2 cm compared to the 0.5-cm azimuthal extent used for ground-truth simulation in Part I [23]. The simulations here have $N_i$ = 400 computational nodes, with a time step of $\Delta t = 5 \times 10^{-12} s$. The total simulated time for all the simulations is 10 $\mu s$. The collected data from the simulations are the relevant plasma properties averaged over every 500 timesteps (2.5 ns). The initial 2 $\mu s$ of the simulation data during which the system exhibits highly nonlinear transient behaviors is discarded.

Along the azimuthal direction, a periodic boundary condition is applied to the particles motion as well as to the electric potential in order to mimic the periodicity along the azimuth. While the electric potential is resolved only along the azimuthal direction (1D problem setup), the particles' trajectories are traced along all the three dimensions. A finite extent of 1 cm is assumed along the axial direction to limit the growth of the energy of the system [39]. The particles crossing an axial boundary of the domain are resampled from their initial distributions and reintroduced into the domain from the opposite axial boundary. The initialization conditions of the simulations are identical to those presented in Part I [23].

We carried out the PIC simulations over a wide range of electric fields, $E_x \in$ {10, 12.5, 15, 17.5, 20, 22.5, 25, 27.5, 30, 32.5, 35, 37.5, 40} $kVm^{-1}$, at a fixed magnetic field of $B_y$ = 20 mT. Similarly, multiple simulations were performed for various magnetic field values, $B_y \in$ {10, 12.5, 15, 17.5, 20, 22.5, 25, 27.5, 30, 32.5, 35, 37.5, 40} mT, with the electric field being fixed at $E_x = 20\ kVm^{-1}$. The data corresponding to these simulations were used to train and test both parametric and ensemble Phi Method ROMs.

The aim of the present test case is to assess the Phi Method's ability to re-discover from the simulations' data the parametric PDE given by Eq. 14. This equation describes the spatiotemporal evolution of the electrons' axial flux density ($J_{ex}$)

$$-qB_y J_{ex} = m_e \frac{\partial(n_e V_{d,ez})}{\partial t} + m_e \frac{\partial(n_e V_{d,ez}^2)}{\partial z} - qn_e E_z + \frac{\partial \Pi_{zz}}{\partial z}. \qquad \text{(Eq. 14)}$$

In Eq. 14, $q$ is the elementary charge, and $J_{ex} = n_e V_{d,ex}$, where $n_e$ and $V_{d,ex}$ are the electrons' number density and axial drift velocity, respectively. Also, $m_e$ is the electron mass, $E_z$ is the azimuthal electric field component. The terms on the right-hand side of Eq. 14 represent different force terms that contribute to the electrons' axial flux density. The first and the second terms represent, respectively, the contributions of the temporal inertia and the convective inertia. The third term captures the contribution of the azimuthal instabilities [37], and the fourth term corresponds to the pressure, with $\Pi_{zz} = K n_e T_{ez}$, where $K$ is the Boltzmann constant, and $T_{ez}$ is the azimuthal electron temperature.

According to Eq. 14, we have defined the library terms for Phi Method to be $n_e V_{d,ez}$, $n_e E_z$, $n_e T_{ez}$ and $n_e V_{d,ez}^2$. Although the underlying physics of this plasma test case and, hence, the distribution of the plasma properties in general depends on both the $E_x$ and the $B_y$, the $E_x$ does not appear explicitly in Eq. 14. To capture the explicit parametric dependency on the $B_y$, nonetheless, we represent the term $B_y J_{ex}$ on left-hand side of Eq. 14 as a variable for which we would like to develop a ROM using Phi Method. In this sense, once the ROM is trained, to obtain the $J_{ex}$ at a certain value of the $B_y$, we first evaluate the $B_y J_{ex}$ term from the model and then divide by the $B_y$ value.

We have developed two "standard" parametric Phi Method models with slightly different libraries. The first model, called "parametric model 1 (PM1)", is described by the relation in Eq. 15

$$\left[(B_y J_{ex})_i\right]_k = \left[\begin{bmatrix}(n_e V_{d,ez})_{i-1}\\(n_e V_{d,ez})_i\\(n_e V_{d,ez})_{i+1}\end{bmatrix}^T \begin{bmatrix}(n_e E_z)_{i-1}\\(n_e E_z)_i\\(n_e E_z)_{i+1}\end{bmatrix}^T \begin{bmatrix}(n_e T_{ez})_{i-1}\\(n_e T_{ez})_i\\(n_e T_{ez})_{i+1}\end{bmatrix}^T \begin{bmatrix}(n_e V_{d,ez}^2)_{i-1}\\(n_e V_{d,ez}^2)_i\\(n_e V_{d,ez}^2)_{i+1}\end{bmatrix}^T\right]_k \Phi_{\text{PM1}}. \qquad \text{(Eq. 15)}$$

It is underlined from Eq. 15 that, for PM1, the left-hand side of the equation and the library on the right-hand side are both at the same time step $k$, which is in line with Eq. 14. This means that the derived relation from PM1 ROM establishes the correlations that exist at each time step among the specified terms in the model.

Whereas PM1 includes the term $n_e V_{d,ez}$, it does not account for the time derivative of this quantity, which does appear in Eq. 14. To address this, we include the term $n_e V_{d,ez}$ at both the current time step ($k$) and at the preceding time step ($k-1$). This modification yields a variant of PM1, denoted as "parametric model 2, PM2", and given by Eq. 16

$$\left[(B_y J_{ex})_i\right]_k = \left[\begin{bmatrix}(n_e V_{d,ez})_{i-1}\\(n_e V_{d,ez})_i\\(n_e V_{d,ez})_{i+1}\end{bmatrix}^T_k \begin{bmatrix}(n_e V_{d,ez})_{i-1}\\(n_e V_{d,ez})_i\\(n_e V_{d,ez})_{i+1}\end{bmatrix}^T_{k-1} \begin{bmatrix}(n_e E_z)_{i-1}\\(n_e E_z)_i\\(n_e E_z)_{i+1}\end{bmatrix}^T_k \begin{bmatrix}(n_e T_{ez})_{i-1}\\(n_e T_{ez})_i\\(n_e T_{ez})_{i+1}\end{bmatrix}^T_k \begin{bmatrix}(n_e V_{d,ez}^2)_{i-1}\\(n_e V_{d,ez}^2)_i\\(n_e V_{d,ez}^2)_{i+1}\end{bmatrix}^T_k\right] \Phi_{\text{PM2}}. \qquad \text{(Eq. 16)}$$

In Part I [23], we assessed the sensitivity of the Phi Method ROMs' predictions to the size of the library for the present plasma test case. We observed that the inclusion of additional terms in the library does not significantly affect the model's predictive capability as long as the essential terms are included in the library of candidate terms. However, the presence of unnecessary terms can deteriorate the interpretability of the Phi Method model because the model will no longer be sparse. The inclusion of extra terms can also alter the values of the coefficients found for the essential terms. Moreover, a sparse model is more likely to be generalizable to various scenarios beyond the training dataset. In any case, it is also noted that when an expert knowledge might be lacking and, hence, it would not be possible to identify the relevant (essential) dynamical terms a-priori, the library size will inevitably need to be large. The expansion of the library in such cases ensures that sufficient flexibility is embedded within the model so that it will be able to effectively capture the correlations present in the data. The lack of model sparsity in the cases of large library terms can then be addressed by using sparsity-promoting techniques that aim to optimize the coefficient matrix $\Phi$ in the sparsest possible way so as to effectively eliminate the non-essential terms [40][41].

Considering the above arguments, for the sake of the demonstrations in this paper within the plasma discharge test case, we chose to include in the Phi Method's libraries the minimum set of terms which was needed to represent the dynamics, leaving the incorporation of the sparse optimization for the future work.

The parametric Phi Method models PM1 and PM2 were trained on a dataset comprised by the simulation data corresponding to the parameters set of $E_{x,train} \in \{15, 20, 25, 30, 35\}\ kVm^{-1}$ and $B_{y,train} \in \{15, 20, 25, 30, 35\}$ mT. The remaining data from the simulations with the other values of the $E_x$ and the $B_y$ were reserved to be compared against the ROMs' predictions for testing purposes.

For the plasma discharge test case, ensemble Phi Method ROMs were also developed. Two ensemble models with the libraries corresponding to those for the PM1 and the PM2 were trained on the exact training dataset as that for the "standard" parametric ROMs. The ensemble ROMs were derived using the approach detailed in Section 2.2. The coefficients matrices (operators) associated with the two ensemble ROMs are denoted as $\Phi_{EM1}$ and $\Phi_{EM2}$.

Before moving on to the next subsection, we would highlight that the parametric OPT-DMD was found to not be a suitable choice for the plasma test case due to the strong variations observed for the system's behavior (dynamics) across the parameter space. As a result, the parametric OPT-DMD method failed to provide reasonable interpolations of the modes between the parameters. Consequently, the following subsection exclusively presents the results from Phi Method.

### 3.2.2. Results

Once the Phi Method models were trained, the derived data-driven relations were used to predict the spatiotemporal evolution of the $J_{ex}$ over the unseen (test) parameter set. The predictions of the PM1 and PM2 as were defined by Eqs. 15 and 16, respectively, exhibited indistinguishable similarity. Moreover, PM1 predictions were found to be closely representative of the predictions of the ensemble model EM1 as well. This will be evidenced later in this subsection. Therefore, to avoid redundancy in presenting the results and for the clarity of discussions, Figure 7 to Figure 9 present the predictions of the PM1 only.

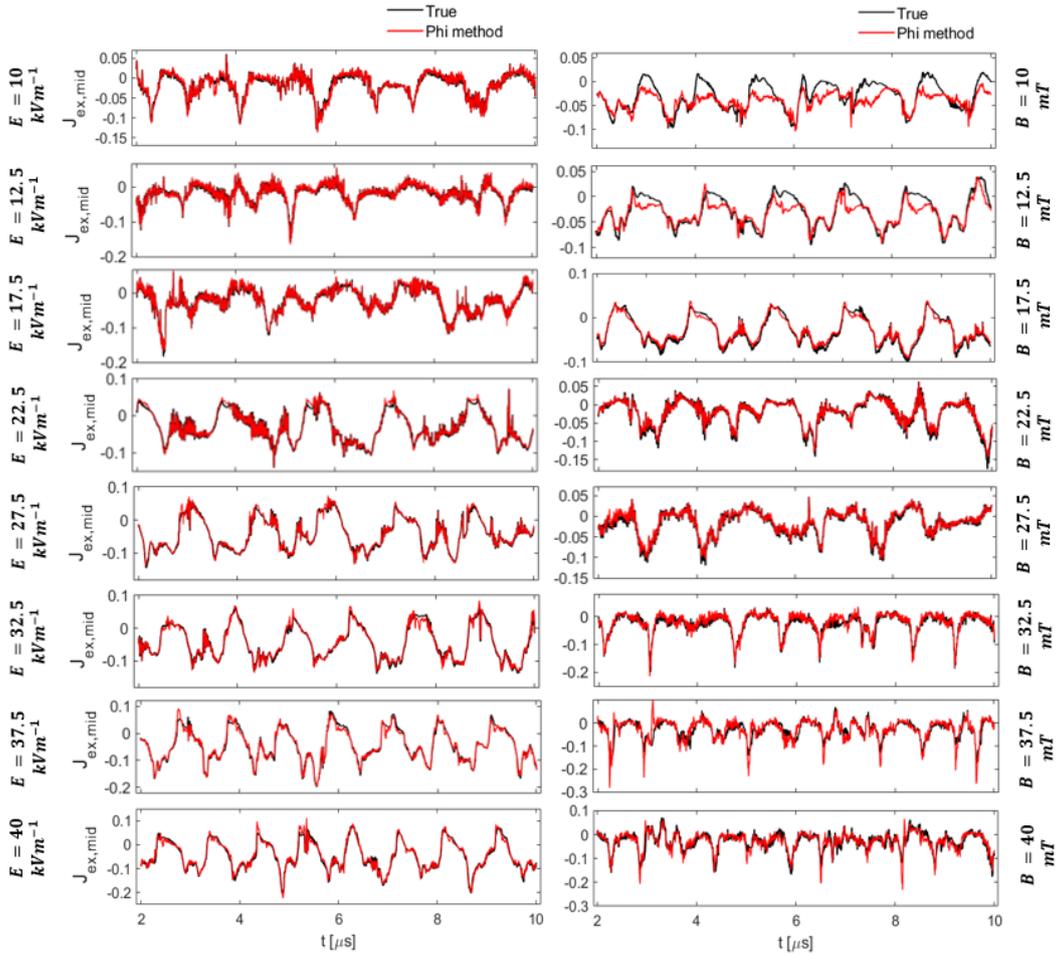

Figure 7: Predictions of the parametric Phi Method ROM (PM1) for the test case 2 compared against the ground-truth data; time evolutions of the local value of the normalized electrons' axial flux density ($J_{ex}$) at the mid-location within the simulation domain for various test values of the axial electric field ($E_x$, **left column**) and for various test values of the radial magnetic field ($B_y$, **right column**).

The predicted time evolutions of the local $J_{ex}$ value at the center of the domain for various values of the applied $E_x$ and $B_y$ belonging to the test parameter set are provided in Figure 7. The predicted signals are compared against the ground-truth data from the PIC simulations. Additionally, in Figure 8 and Figure 9, the complete spatiotemporal maps of the $J_{ex}$ as predicted by the PM1 are compared against the "true" 2D maps from the PIC simulations for the various test values of the $E_x$ and the $B_y$, respectively.

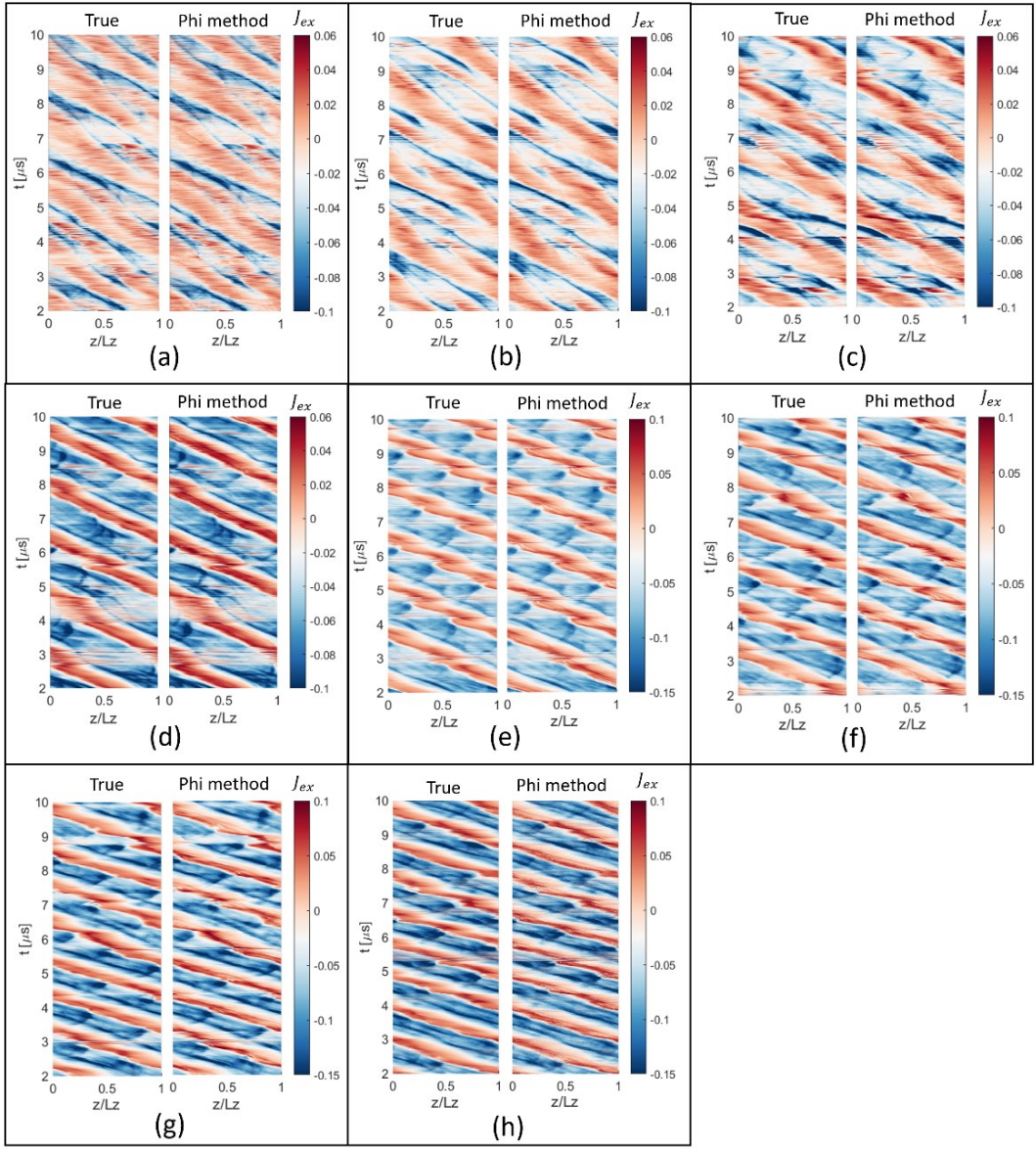

Figure 8: Predictions of the parametric Phi Method ROM (PM1) for the test case 2 compared against the ground-truth data; spatiotemporal maps of the normalized electrons' axial flux density ($J_{ex}$) for the test values of the axial electric field ($E_x$): (a) $E_x = 10\ kVm^{-1}$, (b) $E_x = 12.5\ kVm^{-1}$, (c) $E_x = 17.5\ kVm^{-1}$, (d) $E_x = 22.5\ kVm^{-1}$, (e) $E_x = 27.5\ kVm^{-1}$, (f) $E_x = 32.5\ kVm^{-1}$, (g) $E_x = 37.5\ kVm^{-1}$, and (h) $E_x = 40\ kVm^{-1}$.

Looking at Figure 7 to Figure 9, it is evident that the parametric ROM is able to predict the time evolution of the $J_{ex}$ as a function of the plasma properties included in the Phi Method's library. In addition, the parametric data-driven relationship recovered between the target quantity – $J_{ex}$ – and the candidate terms in the library is applicable to a wide parameter space, which emphasizes the generalizability of the model.

In Figure 7 to Figure 9, the eight extreme cases corresponding to the lowest and the highest values of the applied electric and magnetic fields, namely, $E_x$ = 10, 12.5, 37.5, and 40 $kVm^{-1}$, and $B_y$ = 10, 12.5, 37.5, and 40 mT, fall outside the range of the parameter space over which PM1 had been trained. These cases, thus, represent extrapolation scenarios. In this regard, toward the lowest and the highest magnetic field values, the PM1's accuracy degrades, with the deviations being mainly noticeable around the signals' peaks and dips. However, the predictions are still reasonably aligned with the ground truth. For the extreme electric field cases, the accuracy of the PM1 predictions remains at the same level as that in the intermediate-parameter regimes.

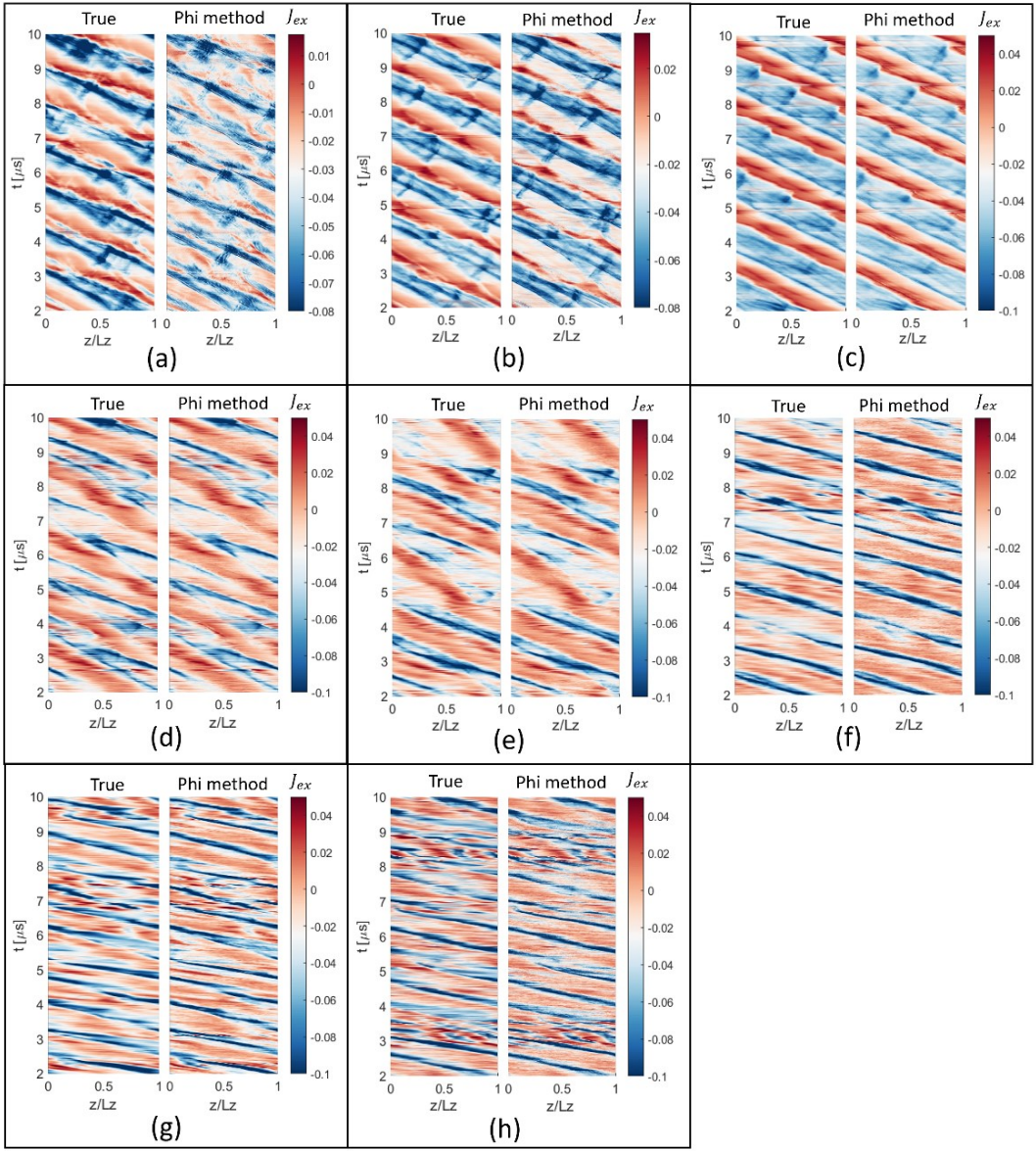

Figure 9: Predictions of the parametric Phi Method ROM (PM1) for the test case 2 compared against the ground-truth data; spatiotemporal maps of the normalized electrons' axial flux density ($J_{ex}$) for the test values of the radial magnetic field ($B_y$): (a) $B_y = 10$ mT, (b) $B_y = 12.5$ mT, (c) $B_y = 17.5$ mT, (d) $B_y = 22.5$ mT, (e) $B_y = 27.5$ mT, (f) $B_y = 32.5$ mT, (g) $B_y = 37.5$ mT, and, (h) $B_y = 40$ mT.

To have an overall assessment of the variations in the ROMs' accuracy across the investigated parameter space, the spatiotemporally averaged predicted values of the $J_{ex}$ from the PM1 and the EM1 models are plotted in Figure 10 vs the $E_x$ and the $B_y$ values. Figure 10(a) and (c) show the variations vs the $E_x$ and the $B_y$ for the PM1. Figure 10(b) and (d) show the same variations for the EM1. The corresponding "true" $J_{ex}$ values from the PIC simulations are also shown in these plots as red dots. We have superimposed on the plots of Figure 10 the spatiotemporally averaged losses of the $J_{ex}$ predictions with respect to the true values from the simulations.

The plots in Figure 10 indicate that, across various values of the $E_x$, the loss factors associated with the PM1 and the EM1 models remain almost constant, whereas they exhibit notable variations across the different $B_y$ values. The PM1's and the EM1's loss increases toward the two bounds of the investigated $B_y$ range.

Comparing the plots (b) and (d) in Figure 10 against plots (a) and (c), we observe that the predictions' accuracy of the EM1 is quite comparable to that of the PM1 overall. Accordingly, we can conclude that, for this 1D plasma discharge test case, either training a single model over a large dataset or combining multiple models trained over subsets of the data lead to nearly equivalent ROMs in terms of their predictive performance.

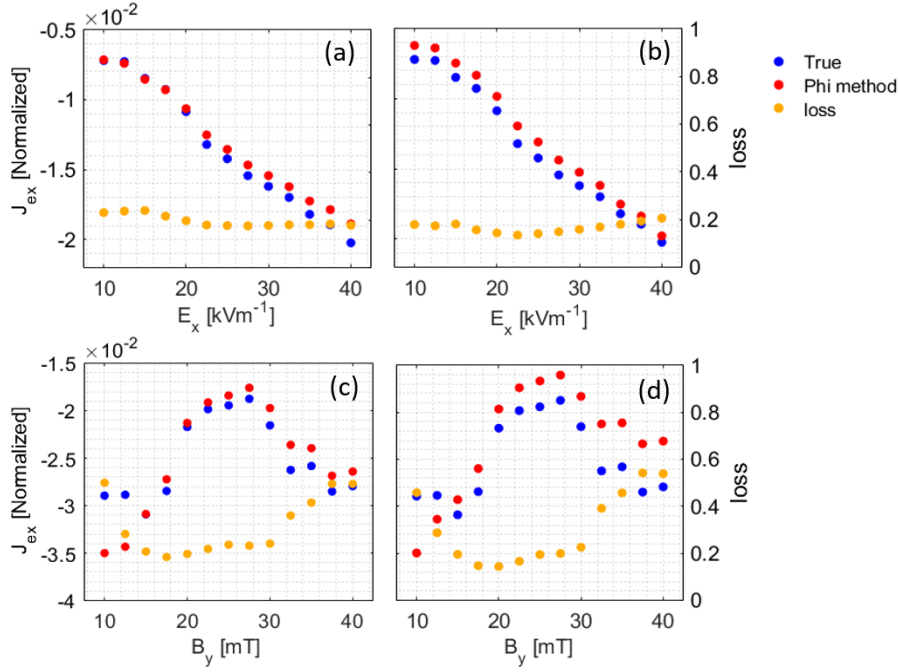

Figure 10: Predictions of the PM1 (**left column**) and the EM1 (**right column**) for the test case 2; spatially and temporally averaged normalized electrons' axial flux density ($J_{ex}$) values from the ROMs (predictions) and the simulations (ground-truth) against: (a) and (b) the various electric field values, (c) and (d) the various magnetic field values. The associated predictions' loss factors are also plotted.

*3.2.2.1. Characteristics and interpretability of the coefficients matrices for the parametric and the ensemble Phi Method ROMs*

Here, we examine closely the obtained $\Phi$ matrices (operators) corresponding to the different DD ROMs discussed so far, namely, the PM1 and the PM2 ("standard" parametric Phi Method ROMs), and EM1 and EM2 (ensemble Phi Method ROMs). We recall from subsection 3.2.1 that $\Phi_{PM1}$ and $\Phi_{EM1} \in R^{12 \times 1}$, whereas $\Phi_{PM2}, \Phi_{EM2} \in R^{15 \times 1}$. For the demonstration purposes in this subsection, these Phi method operators are rearranged into matrices with the dimensions of $3 \times 4$ and $3 \times 5$ for the PM1/EM1 and the PM2/EM2, respectively. The rearranged $\Phi$ matrices are illustrated in Figure 11.

It is observed that the $\Phi$ matrices of the four models exhibit notable similarity in terms of the relative significance of their respective entries. In particular, the variations of the coefficients' values along the horizontal axis, which shows the relative importance of the dynamical terms, present a common trend among the matrices of the various ROMs.

Along the vertical axis, i.e., for each term, the relative magnitudes of the coefficients represent the optimal discretization stencil learned by Phi Method for that particular term. In this regard, the $\Phi_{EM1}$ and the $\Phi_{EM2}$ matrices (Figure 11(b) and (d)) are seen to have identified slightly different stencils for certain terms compared to the parametric ROMs (Figure 11(a) and (c)). The differences are the most distinguishable for the $n_e T_{ez}$ and $n_e V_{d,ez}$ terms.

Furthermore, the coefficients associated with the term $(n_e V_{d,ez})_{k-1}$ in PM2/EM2, which is absent in PM1/EM1, are vanishingly small. The similarity of the magnitudes of the coefficients shared among the various models suggests that the inclusion of the additional term $(n_e V_{d,ez})_{k-1}$ does not have a notable impact on the Phi Method ROM.

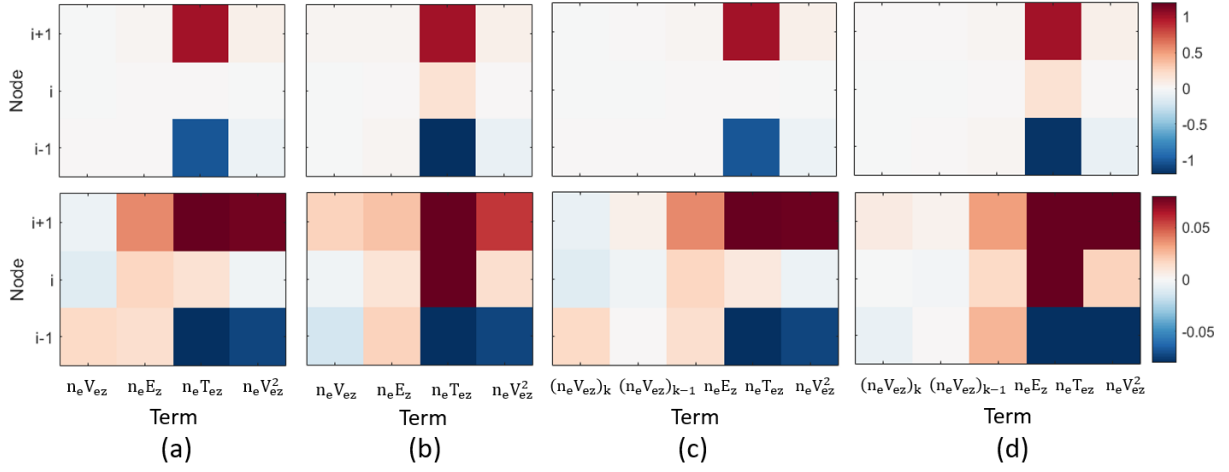

Figure 11: Rearranged normalized representations of the coefficients matrices ($\Phi$) corresponding to the parametric Phi Method ROMs (PM1 and PM2) and the ensemble Phi Method ROMs (EM1 and EM2) for the test case 2; (a) $\Phi_{PM1}$, (b) $\Phi_{EM1}$, (c) $\Phi_{PM2}$, (d) $\Phi_{EM2}$. The bottom row plots are the rescaled visualizations of the top row plots.

Figure 12 presents the normalized standard deviations of the coefficients of the library terms across the individual models that yielded the ensemble Phi Method ROMs. The standard deviation of each coefficient is normalized with respect to the magnitude of the ensemble mean of the corresponding coefficient over the individual models.

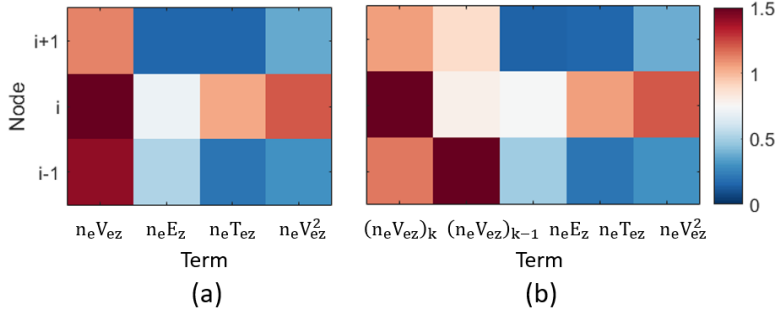

Figure 12: Normalized standard deviations of the coefficients of the library terms across the individual models constituting the ensemble ROMs; (a) standard deviations across the individual models of EM1, (b) standard deviations across the individual models of EM2. The normalization is performed by the absolute value of the mean coefficients matrices ($\Phi_{EM1}$ and $\Phi_{EM2}$).

From Figure 12, the largest variability (standard deviation) among the coefficients of the matrices of the individual models is observed for the coefficients of the $n_e V_{d,ez}$ term. The rest of the terms show lower levels of variability in their coefficients' values over the individual models, especially for the neighboring nodes of $i-1$ and $i+1$. This analysis shows that the coefficients associated with the more dominant terms are determined with a higher level of confidence. Whereas, the determination of the coefficients related to terms that contribute minimally to the dynamics involves the highest degree of uncertainty. In the present test case, the least contributing terms are the $n_e V_{d,ez}$ for EM1 and the $(n_e V_{d,ez})_k$ and $(n_e V_{d,ez})_{k-1}$ for EM2.

As the last discussion, we aim to compare the data-driven coefficients of the Phi Method ROMs against the analytical counterparts from the discretization of Eq. 14. To this end, we rewrite Eq. 14 in a discretized form using the finite differencing scheme

$$-qn_{e,max}B_{max}V_{d,ex_{max}}(\hat{n}_e \hat{B} \hat{V}_{d,ex}) = \frac{m_e}{\Delta t} n_{e,max} V_{d,ez_{max}} \Delta_t(\hat{n}_e \hat{V}_{d,ez}) - qn_{e,max}E_{z,max}(\hat{n}_e \hat{E}_z) + \frac{K}{2\Delta z} n_{e,max} T_{ez,max} \Delta_z(\hat{n}_e \hat{T}_{ez}) + \frac{m_e}{2\Delta z} n_{e,max} V_{d,ez_{max}}^2 \Delta_z(\hat{n}_e \hat{V}_{d,ez}^2)$$

(Eq. 17)

In Eq. 17, $\Delta_t$ denotes first-order-accurate temporal discretization of the time derivative (i.e., $\Delta_t(\hat{n}_e \hat{V}_{d,ez}) = (\hat{n}_e \hat{V}_{d,ez})_k - (\hat{n}_e \hat{V}_{d,ez})_{k-1}$). $\Delta_z$ is the second-order-accurate discretized spatial derivative (e.g., $\Delta_z(\hat{n}_e \hat{T}_{ez}) = \hat{n}_e \hat{T}_{ez}|_{i+1} - \hat{n}_e \hat{T}_{ez}|_{i-1}$). $\Delta z$ represents the size of the cells used for the discretization along the z direction, and $\Delta t$ denotes the timestep between two consecutive data snapshots. Any quantity labeled with the subscript "*max*" represents the spatiotemporal maximum of that quantity across the entire dataset. The quantities marked with the caret symbol (^) on the top indicate the normalized properties with respect to their maximum value. Rearranging

Eq. 17 as in Eq. 18 results in a theoretical relation among the normalized properties, which is equivalent to the data-driven relation from the Phi Method models

$$\hat{B}\hat{J}_{ex} = C_1 \Delta_t(\hat{n}_e \hat{V}d_{ez}) + C_2(\hat{n}_e \hat{E}_z) + C_3 \Delta_z(\hat{n}_e \hat{T}_{ez}) + C_4 \Delta_z(\hat{n}_e \hat{V}^2_{d,ez}),$$ (Eq. 18)

The coefficients in Eq. 18 are defined as

$$C_1 = -\frac{m_e V_{d,ez_{max}}}{qB_{max}V_{d,ex_{max}}\Delta t},$$ (Eq. 19)

$$C_2 = \frac{E_{z,max}}{B_{max}V_{d,ex_{max}}},$$ (Eq. 20)

$$C_3 = -\frac{KT_{ez,max}}{2qB_{max}V_{d,ex_{max}}\Delta z},$$ (Eq. 21)

$$C_4 = -\frac{m_e V^2_{d,ez_{max}}}{2qB_{max}V_{d,ex_{max}}\Delta z}.$$ (Eq. 22)

In order to directly compare the above theoretical coefficients with the Phi Method's learned coefficients, we needed to derive the equivalent Phi Method's approximations of the theoretical coefficients $C_1$, $C_2$, $C_3$, and $C_4$ from the $\Phi$ matrices. To approximate the coefficients of the terms that involve spatial derivatives, i.e., $\Delta_z(\hat{n}_e \hat{T}_{ez})$ and $\Delta_z(\hat{n}_e \hat{V}^2_{d,ez})$, we computed the average of the coefficients of the nodes $i-1$ and $i+1$. For the terms not involving derivatives such as $\hat{n}_e \hat{E}_z$, we considered the average of the coefficients over the three neighboring nodes of $i-1$, $i$, and $i+1$ as a representative of the terms' coefficient. For the temporal derivative term, $\Delta_t(\hat{n}_e \hat{V}_{d,ez})$ that was included in the library of PM2/EM2, the respective coefficient was approximated as the average difference between the coefficients of $(\hat{n}_e \hat{V}_{d,ez})_k$ and $(\hat{n}_e \hat{V}_{d,ez})_{k-1}$ terms across all the three neighboring nodes $i-1$, $i$, and $i+1$.

According to the above paragraph, the relations between the approximated coefficients from PM1/EM1 and PM2/EM2 (denoted by superscript 1 and 2, respectively) and the entries of the respective rearranged matrices of the models ($\Phi^1$ for PM1/EM1 and $\Phi^2$ for PM2/EM2) are defined as

$$C_1^2 \approx \frac{1}{3}\left(\sum_{i=1}^{3}(\Phi^2_{i,1} - \Phi^2_{i,2})\right),$$ (Eq. 23)

$$C_2^1 \approx \frac{1}{3}\left(\sum_{i=1}^{3}(\Phi^1_{i,2})\right), \quad C_2^2 \approx \frac{1}{3}\left(\sum_{i=1}^{3}(\Phi^2_{i,3})\right),$$ (Eq. 24)

$$C_3^1 \approx \frac{1}{2}(\Phi^1_{1,3} + \Phi^1_{3,3}), \quad C_3^2 \approx \frac{1}{2}(\Phi^2_{1,4} + \Phi^2_{3,4}),$$ (Eq. 25)

$$C_4^1 \approx \frac{1}{2}(\Phi^1_{1,4} + \Phi^1_{3,4}), \quad C_4^2 \approx \frac{1}{2}(\Phi^2_{1,5} + \Phi^2_{3,5}).$$ (Eq. 26)

In the above equations, $\Phi_{i,j}$ indicates the entry at the $i$-th row and the $j$-th column of the rearranged $\Phi$ matrix for PM1/EM1 or PM2/EM2.

In Figure 13(a) and (b), the evaluations of the coefficients $C_1$ to $C_4$ from PM1, PM2, EM1, and EM2 are compared against the corresponding analytical values of the same coefficients. Moreover, in Figure 13(c) and (d), the errors defined as the normalized differences between the Phi Method coefficients' values and the corresponding analytical ones are plotted. The normalization factor has been the analytical values of the coefficients.

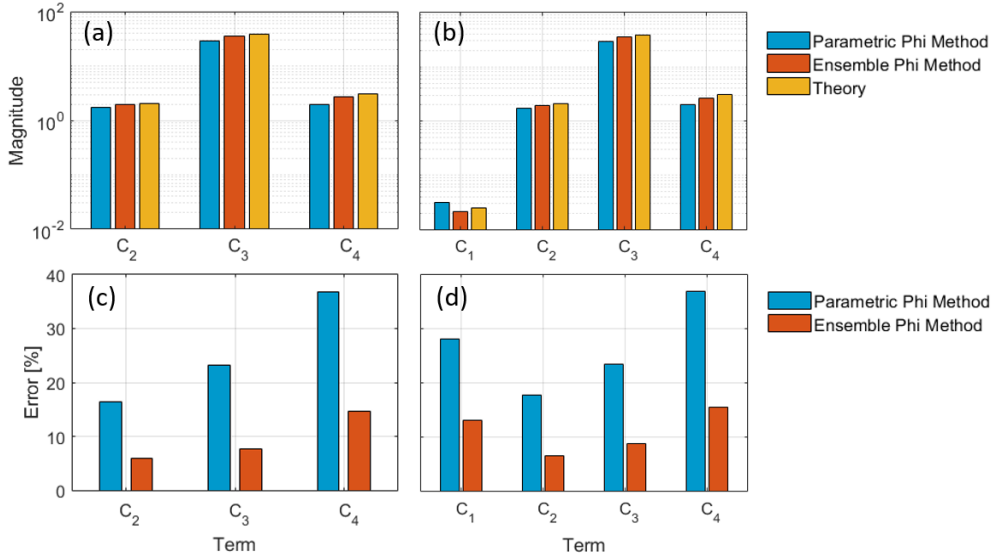

Figure 13: Comparison of the learned coefficients of the terms in Eq. 18 from the parametric and the ensemble Phi Method against the respective analytical values of the coefficients for the plasma discharge test case. Plots (a) and (b) show the magnitudes of the coefficients for the PM1/EM1 and the PM2/EM2 ROMs, respectively. Plots (c) and (d) show, respectively, the normalized error of the obtained coefficients relative to the analytical values for the PM1/EM1 and the PM2/EM2 ROMs.

The plots in Figure 13 indicate that, first of all, all the studied Phi Method ROM variants have identified the coefficients $C_1$ to $C_4$ within a 40 % range of the corresponding analytical values. It shall be noted that, while we have termed the difference between the analytical and the learned coefficients as "error", this does not necessarily imply an inherent error in the model. In fact, in order to derive the analytical coefficients themselves as well as to obtain the approximations of the equivalent coefficients from the $\Phi$ matrices, we assumed the finite difference discretization stencil. However, the Phi Method's premise is that it finds the optimal discretization stencil, which might be different from that of the finite difference method. Furthermore, it is important to note that we used PIC simulations for the dataset generation in this test case, and that the PIC method does not explicitly solve Eq. 14. This means that the finite difference scheme was not part of the solution that led to the generation of the dataset. The absence of the finite difference scheme in the solution process underlying the dataset supports the hypothesis that the finite difference scheme does not necessarily serve as the optimal discretization stencil for the present test case. Nonetheless, we would also mention that the presence of noise in the PIC simulation data can slightly impact the inference of the optimal coefficient matrix, which can also partly account for the observed disparity between the learned and the analytically calculated coefficients.

As the second point, the coefficients obtained from the ensemble models appear to align more closely with the "theory". However, this apparently higher level of agreement might be misleading, considering our above argument regarding the caveat of assuming the finite differencing stencil to translate the coefficients of the $\Phi$ matrices back to the coefficients of Eq. 18. This is especially true for the terms involving spatial derivatives because, during the translation of the learned coefficients based on the finite difference stencil, the contribution of the middle node (node $i$) would be ignored. In this regard, and as an example, we can see from Figure 11(b) and (d) that, while the middle node's coefficient for the $n_e T_{ez}$ term is as large as that on the neighboring nodes, there is no obvious way to include its contribution in the approximation of the $C_3$ coefficient.

Despite the above remark, the improved agreement between the ensemble ROMs' coefficients and the analytical ones can also be, at least partly, due to the statistical benefits of the ensembling approach. In general, the ensembling can average out the biases and errors of the individual models, leading to a more representative estimation of the underlying data patterns. Therefore, despite the equivalent prediction accuracy observed between the "standard" parametric Phi Method ROMs and the ensemble ROMs, the ensembling might have, to certain extent, improved the estimation of the coefficients matrix.

**Section 5: Conclusions**

Following on the introduction of Phi Method in Part I as a novel data-driven local operator-finding approach to enable the discovery of discretized partial differential equations governing the dynamics of physical systems, especially plasma configurations, in this Part II, we presented the application of Phi Method toward the discovery of parametric PDEs – equations that describe the time evolution of systems with parametric dependency of the dynamics. We demonstrated this capability of Phi method in two test cases: (1) a 2D fluid flow around a cylinder with parametric dependence on the Reynolds number, and (2) a 1D Hall-thruster-representative E × B plasma discharge in which the underlying phenomena and the global dynamics are largely dependent on the values of the externally imposed electric and magnetic fields.

We discussed that, to learn a data-driven ROM for the dynamics of a parametric system using Phi Method, two distinct approaches are possible: one approach is that of the "standard" parametric Phi Method, which involves learning a single model trained over a range of values of the parameter(s) associated with the system. The other approach – the "ensemble" Phi Method – consists of developing individual data-driven models for each specific parameter value and then obtaining an aggregate model by averaging over the individually trained models. This aggregate (ensemble) model can then be used for the predictions of the dynamics over the unseen parameter space.

In each test case investigated, the ROMs from the parametric and the ensemble implementations of Phi Method were well able to recover the governing parametric PDE from the data of the simulations performed over a training set of values of the relevant parameter(s) (Reynolds number for the fluid system, and the electric and the magnetic field intensities for the plasma system).

The parametric Phi Method ROMs also presented a remarkable performance in predicting the systems' dynamics over the unseen sets of parameters, confirming the generalizability of the learned dynamics over the parameter space.

In the plasma discharge test case, we showed that the predictive performance of the ROMs from the parametric and the ensemble Phi Method is quite comparable. The Φ matrices obtained from either approach were also very similar, particularly in terms of the identified relative importance of the dynamical terms. The equivalent learned coefficients of the PDE describing the time evolution of the electrons' axial flux density from either of these two approaches compared rather closely with the analytical coefficients obtained from the finite-difference-discretization of the same PDE.

We highlighted, nonetheless, that the ensemble Phi Method approach provides certain benefits over the "standard" parametric Phi Method. In this respect, the ensemble Phi Method provides, on the one hand, statistical insights into the derived ROMs. For instance, we understood from the ensemble Phi Method that the coefficients of the terms that dominantly contribute to the dynamics of the plasma system are determined with a high confidence level (small standard deviations across the individual models). On the other hand, the aggregation technique that underpins the ensemble Phi Method can lower the biases and/or errors of the individual models, hence, improving the reliability and robustness of the resulting ensemble ROM.

In the fluid flow test case, we also applied the parametric extension of the OPT-DMD algorithm (Section 2.1) in order to compare the performance of this POD-based approach in representing the parametric dynamics against the performance of Phi Method. The parametric OPT-DMD provided unsatisfactory predictions of the dynamics over the test parameter set. However, the predictions improved in the case where the algorithm was trained on an expanded dataset corresponding to a larger set of training Reynolds numbers. In contrast, the parametric Phi Method was seen to not require a large dataset representing many samples of the parameter space so as to be able to properly represent the parametric dynamics. This observation highlights a strong merit of Phi Method as a local operator-finding algorithm for parametric dynamics discovery over POD-based approaches like DMD.

We would finally emphasize that, even though, for the adopted test cases in this paper, the parametric dynamics and the relationship among the involved state variables and parameters were known, the great performance of Phi Method in recovering the involved parametric dependencies of the dynamics and the demonstrated generalizability of the associated ROMs have important applied implications. Indeed, the results underline that Phi Method can be reliably applied to the scenarios where the underlying parametric PDE(s) may not be fully or partially known, or the cases where one may be interested in finding correlations between a certain state variable and the quantities of the system that can be readily experimentally accessible.

**Acknowledgments**:

The present research is carried out within the framework of the project "Advanced Space Propulsion for Innovative Realization of space Exploration (ASPIRE)". ASPIRE has received funding from the European Union's Horizon 2020 Research and Innovation Programme under the Grant Agreement No. 101004366. The views expressed herein can in no way be taken as to reflect an official opinion of the Commission of the European Union.

MR, FF, and AK gratefully acknowledge the computational resources and support provided by the Imperial College Research Computing Service (http://doi.org/10.14469/hpc/2232).

**Data Availability Statement**:

The simulation data that support the findings of this study are available from the corresponding author upon reasonable request.

**Appendix: Additional results and analyses**

**A. Supplementary results for the test case 1**

In this appendix, supplementary results are provided for the test case 1 from the parametric OPT-DMD and the parametric Phi Method ROMs. The model definitions follow exactly what was described in subsection 3.1.1. The only difference here is that a larger dataset, i.e., DS2, is used for the training of the models. DS2 corresponds to more samples of the parameter space associated with the fluid system. As a reminder, the training set of DS2 includes the simulation data for $Re_{train} \in \{100, 150, 200, 250, 300, 350, 400, 450, 500\}$. The ROMs are tested on the set of Reynolds numbers corresponding to $Re_{test} \in \{175, 275, 425\}$.

Figure 14 shows the predicted time evolutions of the $\Omega_{mean}$ and the $\Omega_{mid}$ (the spatial average and the mid-domain value of the vorticity, respectively) in comparison against the ground-truth. The figure also presents the time evolutions of the loss factors associated with the models' predictions. Figure 15 compares the predicted and the true snapshots of the vorticity field at three sample time instants. Figure 16 shows a comparison between the true and the interpolated spatial modes corresponding to the first four leading DMD bases across the test $Re$ parameters of DS2.

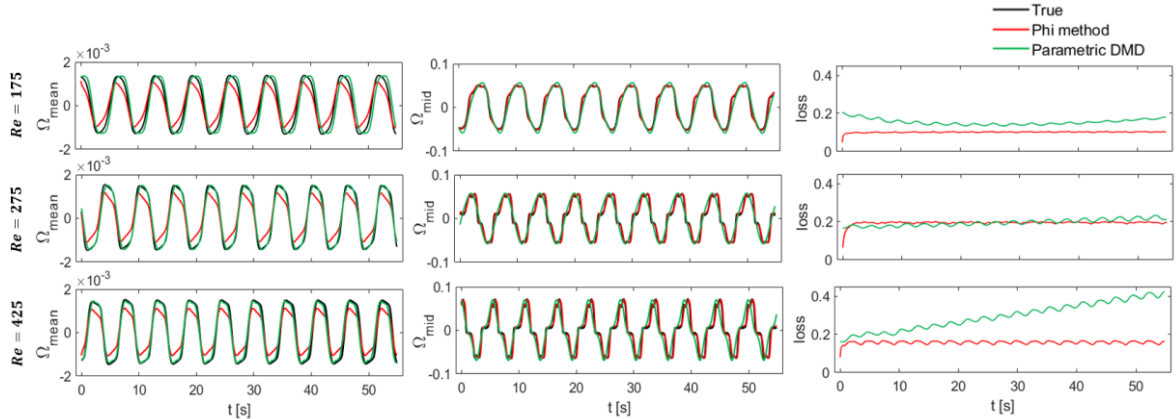

Figure 14: Comparison of the predictions from the parametric Phi Method and the parametric OPT-DMD ROMs against the ground-truth data for the test case 1 across the test Reynolds numbers belonging to DS2; time evolutions of the spatially averaged normalized vorticity (**left column**), and local values of the normalized vorticity at the mid-location within the simulation domain (**middle column**). (**right column**) Time evolutions of the loss factor calculated over the entire domain.

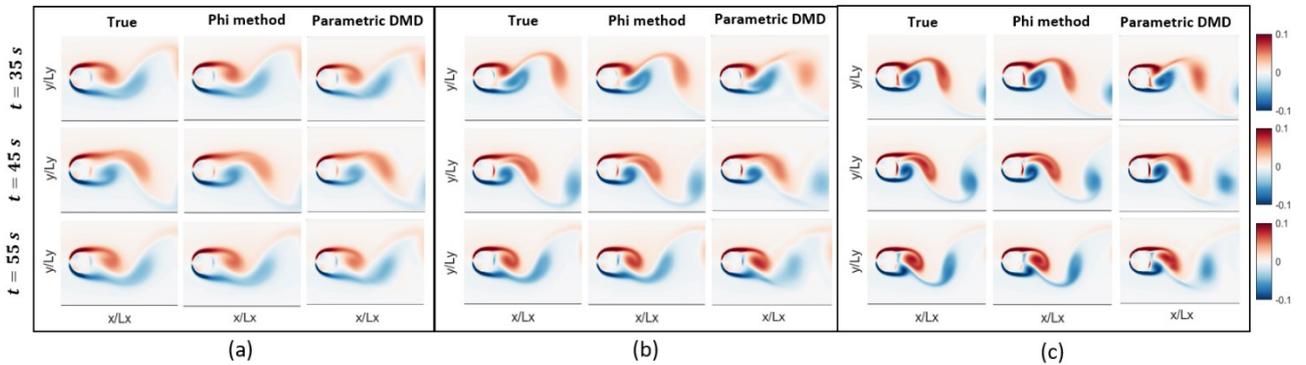

Figure 15: Comparison of the predictions from the parametric Phi Method and the parametric OPT-DMD ROMs against the ground-truth data for the test case 1 across the test Reynolds numbers belonging to DS2; 2D snapshots of the normalized vorticity field at three different time instants for the test parameters of (a) $Re = 175$, (b) $Re = 275$ and (c) $Re = 425$.

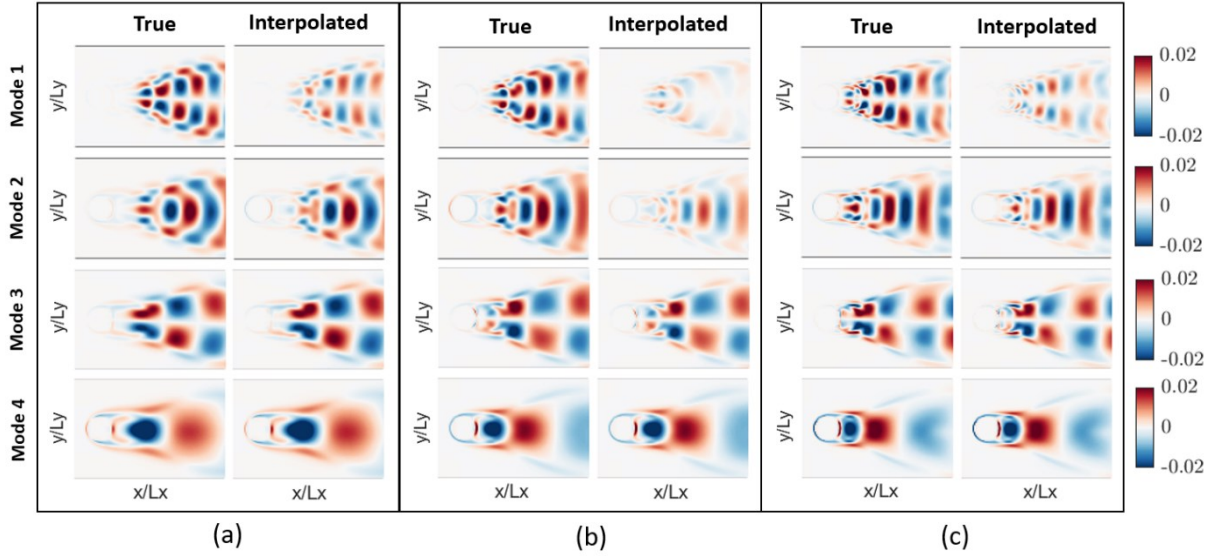

Figure 16: Visualization of the first four interpolated DMD modes associated with the parametric OPT-DMD for the test case 1 over the test Reynolds numbers of DS2. The interpolated modes are compared against the corresponding true DMD modes. (a) $Re = 175$, (b) $Re = 275$ and (c) $Re = 425$.

The above outcomes, together with those presented in subsection 3.1.2, show that a smaller increment between the training parameters that sample the parameter space of the system improves the quality of the parametric OPT-DMD model. A "sufficient" number of parameter samples is case-specific and depends on the extent to which the DMD modes vary within the parameter space of a certain system.

## B. Ensemble Phi Method model for the test case 1

In this appendix, we present the rearranged representations of characteristic matrices corresponding to the ensemble Phi Method model for the fluid flow test case.

Figure 17(a) shows the $\Phi_{Ens}$ matrix (the mean of the $\Phi$ matrices of the individual ROMs). Figure 17(b) illustrates the normalized standard deviations of the coefficients of the library terms across the individual models. Most notably, the $\Phi_{Ens}$ operator of the ensemble ROM is very similar to the $\Phi_{Param}$ associated with the parametric Phi Method ROM (Figure 6) in terms of the coefficients of the terms.

It is noted that the term $\frac{\Omega}{Re}$, which captures the dependency of the dynamics on the $Re$ parameter is not shown in the rearranged matrix representations of Figure 17 because its coefficients were found to be much smaller than the coefficients of the other terms in the Phi Method models' libraries.

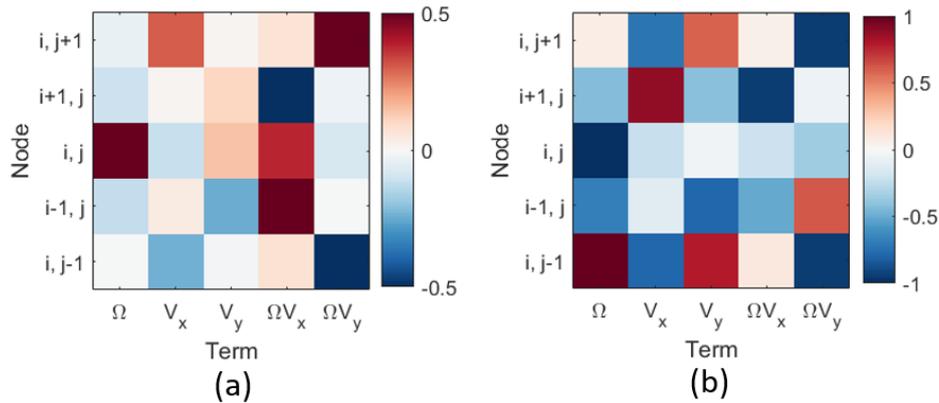

Figure 17: Rearranged normalized representations of the characteristic matrices corresponding to the ensemble Phi Method ROM for the fluid flow test case. (a) $\Phi_{Ens}$ matrix (mean of the $\Phi$ matrices of the individual constituent models), (b) normalized (by the absolute values of the $\Phi_{Ens}$ matrix) standard deviations of the coefficients of the library terms across the individual models. The individual models constituting the ensemble ROM were trained on the dataset 2. The colormap in plot (b) is in logarithmic scale.